\documentclass[final,fleqn,10pt]{siamltex}
\usepackage{url}
\usepackage{fixltx2e}
\usepackage{rotating}
\usepackage{color}
\usepackage{array}
\usepackage{hyperref} 
\usepackage{threeparttable} 
\definecolor{grey}{rgb}{0.95, 0.95, 0.95}
\usepackage[final]{listings}
\usepackage[font=footnotesize]{subfig}
\usepackage{sidecap}
\usepackage{bm}
\usepackage{listings}
\usepackage{graphicx}
\usepackage{icomma}
\usepackage{amsmath}
\usepackage{amssymb}
\usepackage{multirow}
\usepackage{ulem}
\usepackage[usenames,dvipsnames]{xcolor}
\newcommand{\bq}{\begin{equation}}
\newcommand{\eq}{\end{equation}}
\newcommand{\bi}{\begin{itemize}}
\newcommand{\ei}{\end{itemize}}

\newcommand{\FLOP}{\mbox{flop}}
\newcommand{\FLOPS}{\mbox{flops}}

\newcommand{\GBS}{\mbox{GB/s}}

\newcommand{\GFS}{\mbox{GF/s}}

\newcommand{\GHZ}{\mbox{GHz}}

\newcommand{\BYTES}{\mbox{bytes}}

\newcommand{\BITS}{\mbox{bits}}
\newcommand{\BIT}{\mbox{bit}}

\newcommand{\GiB}{\mbox{GiB}}
\newcommand{\MiB}{\mbox{MiB}}
\newcommand{\KiB}{\mbox{KiB}}

\newcommand{\eos}{~.}
\newcommand{\cma}{~,}
\newcommand{\nvidia}{Nvidia}

\newcommand{\SELLCS}{SELL-$C$-$\sigma$}
\newcommand{\cocc}{chunk occupancy}
\renewcommand{\emph}{\textit}
\DeclareCaptionType{copyrightbox}

\begin{document}

\title{A unified sparse matrix data format\\for efficient general sparse matrix-vector multiply on modern processors with wide SIMD units}

%
\author{%
Moritz Kreutzer\footnotemark[1]\footnotetext[1]{Erlangen Regional Computing Center, Friedrich-Alexander-Universit\"at Erlangen-N\"urnberg, D-91058 Erlangen, Germany}\and Georg Hager\footnotemark[1]\and Gerhard Wellein\footnotemark[1]\and Holger Fehske\footnotemark[2]\footnotetext[2]{Institut f\"ur Physik, Ernst-Moritz-Arndt-Universit\"at Greifswald, D-17489 Greifswald, Germany}\and Alan R. Bishop\footnotemark[3]\footnotetext[3]{Theory, Simulation and Computation Directorate, Los Alamos National Laboratory, New Mexico 87545, USA}}

\date{\today}

\maketitle
\begin{abstract}
  Sparse matrix-vector multiplication (spMVM) is the most
  time-consuming kernel in many numerical algorithms and has been
  studied extensively on all modern processor and accelerator
  architectures.  However, the optimal sparse matrix data storage
  format is highly hardware-specific, which could become an obstacle
  when using heterogeneous systems.  Also, it is as yet unclear how
  the wide single instruction multiple data (SIMD) units in current
  multi- and many-core processors should be used most efficiently if
  there is no structure in the sparsity pattern of the matrix.  We suggest \SELLCS,
  a variant of Sliced ELLPACK, as a SIMD-friendly data format which combines
  long-standing ideas from General Purpose Graphics Processing Units (GPGPUs) 
  and vector computer programming. We
  discuss the advantages of \SELLCS\ compared to established formats like
  Compressed Row Storage (CRS) and ELLPACK and show its suitability on a variety of hardware
  platforms (Intel Sandy Bridge, Intel Xeon Phi and \nvidia\ Tesla
  K20) for a wide range of test matrices from different application
  areas. Using appropriate performance models we develop deep insight
  into the data transfer properties of the \SELLCS\ spMVM kernel. \SELLCS\
  comes with two tuning parameters whose performance impact across
  the range of test matrices is studied and for which reasonable
  choices are proposed. This leads to a hardware-independent
  (``catch-all'') sparse matrix format, which achieves very
  high efficiency for all test matrices across all hardware platforms.
\end{abstract}

\section{Introduction and Related Work}
\label{sec:intro}

\subsection{Sparse matrix-vector multiplication on modern hardware}
Many applications in science and engineering are based on
sparse linear algebra. The solution of large eigenvalue problems or
extremely sparse systems of linear equations is a central part of many
numerical algorithms from quantum physics to fluid dynamics to
structural mechanics. The solvers are typically composed of iterative
subspace methods, including advanced preconditioners. At the lowest
level, the multiplication of large sparse matrices with dense vectors
(spMVM) is frequently one of the
most time-consuming building blocks. Thus, the efficient
implementation of this operation is of very high importance.

The spMVM kernel is usually memory-bound for realistic problems on all
modern computer architectures, since its code balance (ratio of main
memory data accesses to executed floating-point operations) is quite
large compared to typical machine balance values (ratio of maximum
memory bandwidth to arithmetic peak performance).  
Additional complications arise because the sparsity pattern of
the matrix, i.e., the position of the non-zero entries, can have
considerable impact on spMVM performance due to indirect access to the
right hand side (RHS) vector; this makes it difficult to understand or
even predict performance via simplistic bandwidth-based modeling.  
And finally, the
sparse matrix storage format has a considerable performance impact and the
optimal choice is known to be very sensitive to the underlying
hardware. Consequently there is nowadays a large variety of sparse
matrix storage formats to choose from.  Some are more suitable for
cache-based standard microprocessors (like CRS), while others yield better performance on vector computers
(like ``Jagged Diagonals Storage'' [JDS]) or on graphics processing
units (like ELLPACK and its variants).

Emerging coprocessors/accelerators like the Intel Xeon Phi or \nvidia\
Tesla GPGPUs are of
special interest for executing spMVM because of their large memory
bandwidth combined with a very high level of on-chip parallelism.
These new compute devices are an integral part of several
supercomputers already today. One may speculate that their 
proliferation will further
increase, making (strongly) heterogeneous compute node architectures
the standard building block of future cluster systems. Thus,
sustainable and modern high-performance parallel software should be
able to utilize both the computing power of accelerators as well as
standard CPUs \emph{in the same system}. As of today, in such a
setting one is forced to deal with multiple sparse storage formats
within the same application code. Hence, it is of broad interest to
establish a single storage format that results in good performance for
all architectures. Since even the current standard microprocessors 
feature SIMD execution or related techniques, such a format
would have to support SIMD parallelism in an optimal way.

It has to be stressed that standard x86-based server microprocessors
are usually so bandwidth-starved even with scalar code (i.e., they
have a low machine balance) that a strongly memory-bound loop kernel
such as spMVM does not benefit very much from SIMD vectorization,
unless the working set is small enough to fit into a cache.  However,
SIMD does make a difference on designs with many very slow cores (such
as the Intel Xeon Phi), and certainly on the massively threaded
GPGPUs. Moreover it can be shown that efficient (i.e.,
SIMD-vectorized) single-core code can yield substantial energy savings
on standard multicore processors by reaching the bandwidth saturation
point with fewer cores \cite{hager:cpe13,sc13lbm}.

\subsection{Related work}
The high relevance of the spMVM operation for many application areas
drives continuous, intense research on efficient spMVM implementations
on all kinds of potential compute devices. This is why we here only briefly
review relevant work in the context of establishing a single matrix
data format for processor architectures used in modern supercomputers.

On cache-based CPUs the CRS format,
as presented by Barrett et al. \cite{barrett94}, still sets the
standard if no regular matrix substructures can be exploited.
Further work, especially on auto-tuning the performance of spMVM kernels
on multi-core CPUs, has been done by Williams et al. \cite{williams07}.
A detailed study of CRS performance characteristics on CPU
architectures has been presented, e.g., by Goumas et al.
\cite{goumas08}.

A first comprehensive analysis of spMVM performance for GPGPUs can be
found in Bell and Garland \cite{bellgarland09}, who adopted the
ELLPACK sparse matrix format which had been used on classic vector
computers by Kincaid et al. \cite{kincaid89} long before. Further
research on this topic towards auto-tuning has been conducted by 
Choi et al. \cite{choi10}.  These efforts have inspired a lot of
subsequent work on more efficient data formats for sparse matrices on
GPGPUs \cite{vazquez11,monakov10,dziekonski11,spmvm12-KHW}. A common
finding in those publications is that ELLPACK-like matrix formats
(such as ELLPACK, ELLPACK-R, ELLR-T, Sliced ELLR-T, pJDS) deliver the
best performance for spMVM on GPGPUs.

The recent appearance of the Intel Xeon Phi coprocessor has spawned
intense research activity around the efficient implementation of
numerical kernels, including spMVM, on this
architecture. First work from Saule et al. \cite{saule13} is based on
the (vectorized) CRS format and showed that this format is in general
not suited for Intel Xeon Phi.

Liu et al.~\cite{Liu:2013:ESM:2464996.2465013} have recently published 
a sparse matrix data format for the Intel Xeon Phi that is very similar 
to ours. They have obtained
much better performance than with CRS, but the performance analysis of
the format was focused on the Xeon Phi architecture, and it was not
applied to GPGPUs and standard microprocessors.

Still, portability of these hardware-specific formats across different
platforms remains an open issue. Recently an spMVM framework based on
OpenCL has been introduced ~\cite{Su:2012:CCO:2304576.2304624}, which
allows for code portability but does not provide 
a unified and efficient spMVM data format across  compute
devices of different hardware architecture. 
Such a format is highly desirable in order to address data distribution issues
arising with dynamic load balancing and fault tolerance on heterogeneous 
systems: Re-distributing matrix 
data becomes a lot easier when all devices use the same storage format.
In addition, a unified data format simplifies the definition and
implementation of interfaces of numerical multi-architecture libraries.

\subsection{Contribution}\label{sec:contribution}
This work demonstrates the feasibility of a single storage format for
sparse matrices, which we call ``\SELLCS.'' It builds on Sliced
ELLPACK \cite{monakov10} and delivers competitive performance on a
variety of processor designs that can be found in modern heterogeneous
compute clusters. Note that Sliced ELLPACK has only been used
on GPGPUs up to now.

We examine the CRS and \SELLCS\ formats specifically in terms of their
suitability for SIMD vectorization on current x86 processors with
``Advanced Vector Extensions'' (AVX) and on the Intel
``Many Integrated Core'' (MIC) architecture.  \SELLCS\ shows best performance if the ``chunk
structure'' of the format is chosen in accordance with the relevant
SIMD width $C$, i.e., the width of a SIMD register on x86 and Intel MIC.
On GPGPUs this is the number of threads per warp.
Sorting rows by the number of non-zero entries within a limited
``sorting scope'' $\sigma$ of rows reduces the overhead of
the scheme and improves performance on all architectures
if $\sigma$ is not too large.

In contrast to previous work on Sliced ELLPACK \cite{monakov10}, our
analysis is complemented by a thorough performance modeling approach
which allows us to understand the influence of the two parameters $C$ and
$\sigma$ on the performance and their interaction with basic matrix
properties such as the ``\cocc'' (related to zero
fill-in) and the number of non-zero entries per row.

Our analysis extends the work of Liu et al.~\cite{Liu:2013:ESM:2464996.2465013} 
as it demonstrates that a single SIMD-optimized
data format is appropriate for all current HPC architectures. 

Using the matrices from the ``Williams Group'' in the University of
Florida matrix collection, we finally demonstrate that a single data
format and a SIMD-vectorized or CUDA-parallelized spMVM kernel with
fixed values for $C$ and $\sigma$ shows best or highly competitive
performance on a standard multi-core processor (Intel Xeon Sandy
Bridge), the Intel Xeon Phi accelerator, and the \nvidia\ Kepler K20
GPGPU for most matrix types.

This paper is strictly limited to the single-chip case, and we use
OpenMP threading only. MPI and hybrid MPI+X parallelization (where X
is a threading or accelerator programming model) is left for future
work. Our spMVM formats also assume ``general'' matrices, i.e., we do not
exploit special substructures that would enable optimizations such as
blocking or unrolling. Adding those on top of the \SELLCS\ matrix 
format implementation will be a challenge in itself.

\section{Hardware and test matrices}

\subsection{Hardware characteristics}
\label{sec:hardware}
For the performance evaluation three modern multi- and many-core
architectures have been chosen in order to cover different architectural
concepts which are of importance for current and future compute
devices:
\begin{itemize}
\item The Intel Xeon Sandy Bridge EP system (``Intel SNB'') stands
 for the class of classic cache-based x86 multi-core processors with a
 moderate number of powerful cores, moderate SIMD acceleration and
 still rather high clock frequency. 
\item Trading core complexity and clock speed for core count 
 and wide SIMD parallelism, the Intel Xeon Phi
 (``Intel Phi'') accelerator marks the transition from traditional
 multi-core technology to massively parallel, threaded
 architectures. 
\item The \nvidia\ Kepler (``\nvidia\ K20'') architecture finally
 represents the class of GPGPU accelerators with their extreme level of
 thread parallelism, reduced core and execution complexity, and a
 different memory subsystem design.
\end{itemize}
Relevant specific key features of these compute devices
are summarized in \autoref{tab:architectures} and are briefly 
described below.
  
\begin{table*}[tbp]
\begin{center}
\begin{threeparttable}[t]
\small\centering\renewcommand{\arraystretch}{1.2}
\begin{tabular}{rcccccc@{}}
& Cores & Clock & LLC & Copy BW & Read BW & SIMD width\\
& & (\GHZ) & (\MiB) & (\GBS) & (\GBS) & (\BITS) \\
\hline\hline
Intel Xeon E5-2680 & 8 & 2.7 & 20 & 36 & 43 & 256  \\
Intel Xeon Phi 5110P & 60 & 1.05 & 30 & 152 & 165 & 512 \\
\nvidia\ Tesla K20c & 2496 & 0.71 & 1.25 & 151 & 124 & 2048 \\
\end{tabular}
\end{threeparttable}
\end{center}
\caption{Relevant technical features of the test systems. The last
 level cache (LLC) size is the size of the largest cache on each
 architecture. The achievable main memory bandwidth (BW) was
 determined using an array copy and a read-accumulate benchmark,
 respectively, in order to get sensible baselines for different matrix
 types (see text for details).}
\label{tab:architectures}
\end{table*}

The Intel SNB is a single socket of a standard
two-socket ``Intel Xeon E5-2680'' server. It is based on Intel's
Sandy Bridge-EP microarchitecture and supports the
AVX instruction set, which works on
256-\BIT\ wide SIMD registers.  It has eight cores with up to two
hardware (SMT) threads each, and is equipped with 32\,\KiB\ of L1 and
256\,\KiB\ of L2 cache per core. The shared L3 cache has a total size
of 20\,\MiB. Note that we restrict our measurements to a single
socket to avoid potential interference from the ccNUMA characteristics
of multi-socket shared memory systems.

The ``Intel Xeon Phi 5110P'' is based on Intel's MIC architecture. 
It is a PCIe-based accelerator card
comprising 60 rather simple cores (based on the P54C design, which was launched in 1994) with four hardware (SMT) threads each. The
hardware threading is intended to compensate the deficiencies of the
in-order core architecture.  Each core is extended by a 512-\BIT\ wide
SIMD unit, which can perform up to eight double-precision (or 16
single-precision) floating-point fused multiply-add operations in a
single instruction.  The Intel Xeon Phi has a shared but segmented L2
cache of 30\,\MiB\, with each segment of 512\,\KiB\ being attached to
a single core. The L2 cache design has several known shortcomings;
e.g., if the same cache line is used by multiple cores then each of
them will hold a separate copy in its local L2 segment. This may
reduce the effective L2 cache size for shared-memory parallel codes to
512\,\KiB\ in the worst case. The coprocessor is equipped with
8\,\GiB\ of global GDDR5 memory. ECC memory protection is available
and was turned on for all measurements in this work. Using the SIMD
units through code vectorization is essential to achieve reasonable
performance on this architecture.

The ``\nvidia\ Tesla K20c'' is based on the ``Kepler''
architecture. It has 13 ``Streaming Multiprocessors'' (SMX),
each with 192 single-precision CUDA cores, for a total of
2496 CUDA cores.  Each double-precision unit is shared among three
CUDA cores, for a total of 64 double-precision units per SMX. Within
each multiprocessor, most of the hardware units are driven in a
so-called ``single instruction multiple threads'' (SIMT) manner: A
group of 32 threads, called ``warp,'' executes the same instruction
at a time.  The card has 1280\,\KiB\ of L2 cache and 5\,\GiB\ of
global GDDR5 memory (with ECC memory protection, which was turned on
for all measurements). Best memory performance is achieved if
all threads of a warp access consecutive elements of an array at the
same time (``load coalescing''). 

Code compilation was done with the Intel C Compiler (ICC) 13.1.0 
for the Intel machines, and with the CUDA Toolkit 5.0 for the 
\nvidia\ GPGPU. The Likwid
tools\footnote{\url{http://code.google.com/p/likwid}} were used
for hardware performance counter measurements 
(e.g., memory bandwidth and energy) on the Intel SNB, and 
for controlling thread affinity on the Intel SNB and Phi.

A practical range for the achievable main memory bandwidth on Intel
architectures is typically set by two ``corner case'' microbenchmarks
(see \autoref{tab:architectures}): The ``copy'' benchmark
represents the unfavorable case, while a read-only bandwidth benchmark
(see \autoref{listing:reduction}) sets an upper limit.
\lstset{caption={Reduction benchmark for read-only main memory bandwidth measurement}\label{listing:reduction}}
\begin{lstlisting}[float=tbp]
#pragma omp parallel for reduction(+:sum)
for(i = 0; i < N; ++i) {
    sum += a[i];
}
\end{lstlisting}
Note that ``non-temporal stores,'' e.g., stores that bypass the cache
and avoid the otherwise mandatory cache line write-allocate transfers
on every write miss, were not used for the copy benchmark on either Intel
architecture. Instead, the measured bandwidth available to the loop
kernel was multiplied by $1.5$ to get the actual transfer rate over
the memory interface. This mimics the data transfer properties for
``skinny'' sparse matrices with very few non-zero elements per row.

The slow read-only performance on the \nvidia\ K20
(see \autoref{tab:architectures}) reflects the difficulties with
performing reduction operations on this architecture, even if they
only happen within shared memory; the global reduction is even omitted
in our case.

From an architectural view we put a GPGPU warp on a level with a SIMD
execution unit (see e.g.\cite{Volkov:2008:BGT:1413370.1413402} for a
more detailed discussion). Thus, we assign a SIMD width of $32\cdot
64\,\BITS = 2048\,\BITS$ to the \nvidia\ K20 in
\autoref{tab:architectures}, assuming that each thread of a warp
processes one double precision data item at a time.

\subsection{Benchmark matrices}
\label{sec:matrices}
\renewcommand{\labelenumi}{\textbf{(\alph{enumi})}} 

We conduct the detailed performance analysis of various storage 
formats based on the four matrices RM07R, kkt\_power, Hamrle3, and ML\_Geer
from The University of Florida Sparse Matrix
Collection\footnote{\url{http://www.cise.ufl.edu/research/sparse/matrices}}. 
Their descriptions can be found in \autoref{sect:cornercases}. 
These specific matrices were chosen because they represent corner cases
of matrix characteristics, which crucially influence the efficiency of the
data layout.

The broad applicability of our insights is then validated against the
matrices of the ``Williams group'' from The University of Florida
Sparse Matrix Collection, available for download from
\nvidia\footnote{\url{http://www.nvidia.com/content/NV\_Research/matrices.zip}}.
These matrices have already been used in previous research
\cite{williams07,bellgarland09,choi10} for analyzing the spMVM on
GPGPUs, and thus provide a good basis for comparison.

Note that the four corner cases together with the 14 matrices from the ``Williams group``
in our opinion constitute a sufficiently large set of test matrices in order to show the general
applicability of our insights.

Basic properties such as the dimension $N$, the number of non-zeros
$N_\mathrm{nz}$, the average number of non-zeros per row
$N_\mathrm{nzr}$, and the density (fraction of non-zeros) of all
considered matrices can be found in \autoref{table:matrices}.
The parameter $\beta$ will be introduced in \autoref{sect:sellpit}.
\begin{table*}[tbp]
\small\centering\renewcommand{\arraystretch}{1.2}
\begin{tabular}{@{}rrrrrrrr@{}}
\# & Test case & $N$ & $N_\mathrm{nz}$ & $N_\mathrm{nzr}$ & density & $\beta_{\sigma=1}^{C=16}$ & $\beta_{\sigma=256}^{C=16}$  \\
\hline\hline
1 & RM07R  &  381,689  &  37,464,962  &  98.16  &  2.57e-04  &  0.63  &  0.93 \\
2 & kkt\_power  &  2,063,494  &  14,612,663  &  7.08  &  3.43e-06  &  0.54  &  0.92 \\
3 & Hamrle3  &  1,447,360  &  5,514,242  &  3.81  &  2.63e-06  &  1.00  &  1.00 \\
4 & ML\_Geer  &  1,504,002  &  110,879,972  &  73.72  &  4.90e-05  &  1.00  &  1.00 \\
\hline
5 & pwtk  &  217,918  &  11,634,424  &  53.39  &  2.45e-04  &  0.99  &  1.00 \\
6 & shipsec1  &  140,874  &  7,813,404  &  55.46  &  3.94e-04  &  0.89  &  0.98 \\
7 & consph  &  83,334  &  6,010,480  &  72.13  &  8.65e-04  &  0.94  &  0.97 \\
8 & pdb1HYS  &  36,417  &  4,344,765  &  119.31  &  3.28e-03  &  0.84  &  0.97 \\
9 & cant  &  62,451  &  4,007,383  &  64.17  &  1.03e-03  &  0.90  &  0.98 \\
10 & cop20k\_A  &  121,192  &  2,624,331  &  21.65  &  1.79e-04  &  0.86  &  0.98 \\
11 & rma10  &  46,835  &  2,374,001  &  50.69  &  1.08e-03  &  0.70  &  0.96 \\
12 & mc2depi  &  525,825  &  2,100,225  &  3.99  &  7.60e-06  &  1.00  &  1.00 \\
13 & qcd5\_4  &  49,152  &  1,916,928  &  39.00  &  7.93e-04  &  1.00  &  1.00 \\
14 & mac\_econ\_fwd500  &  206,500  &  1,273,389  &  6.17  &  2.99e-05  &  0.37  &  0.82 \\
15 & scircuit  &  170,998  &  958,936  &  5.61  &  3.28e-05  &  0.49  &  0.83 \\
\multirow{2}{*}{16} & \multirow{2}{*}{rail4284}  &  $4,284 \times$  &  \multirow{2}{*}{11,279,748}  &  \multirow{2}{*}{2,632.99}  &  \multirow{2}{*}{2.41e-03}  &  \multirow{2}{*}{0.28}  &  \multirow{2}{*}{0.73} \\[-3pt]
&  &  $1,092,610$  &  &  &  &  &  \\
17 & dense2  &  2,000  &  4,000,000  &  2,000.00  &  1.00  &  1.00  &  1.00 \\
18 & webbase-1M  &  1,000,005  &  3,105,536  &  3.11  &  3.11e-06  &  0.45  &  0.67 \\
\end{tabular}
\caption{Summary of basic matrix characteristics. If only one
 dimension is given in the $N$ column, the matrix is square. The last
 two columns show the \cocc\ (see
 \autoref{sect:sellpit}) of each matrix without ($\sigma=1$) and with
 sorting ($\sigma=256$) for a chunk size of $C=16$.}
\label{table:matrices}
\end{table*}

\section{Matrix formats and spMVM kernels}

\begin{figure}[tbp]
\centering
\subfloat[Source matrix]{\label{fig:spm}\includegraphics*[height=0.25\textheight]{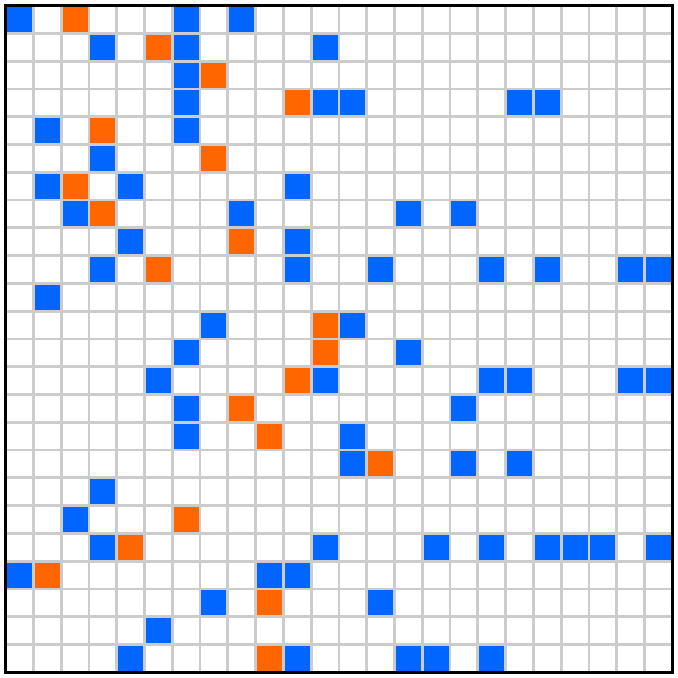}}
\hfill
\subfloat[CRS]{\label{fig:crs}\includegraphics*[height=0.25\textheight]{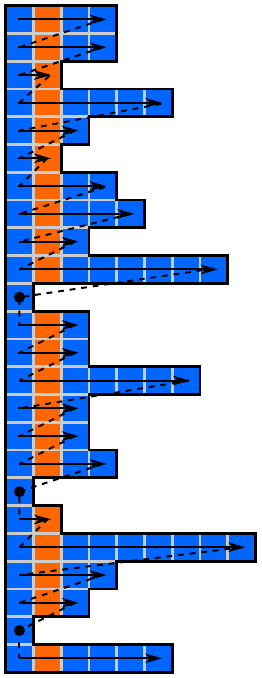}}
\hfill
\subfloat[ELLPACK]{\label{fig:ellpack}\includegraphics*[height=0.25\textheight]{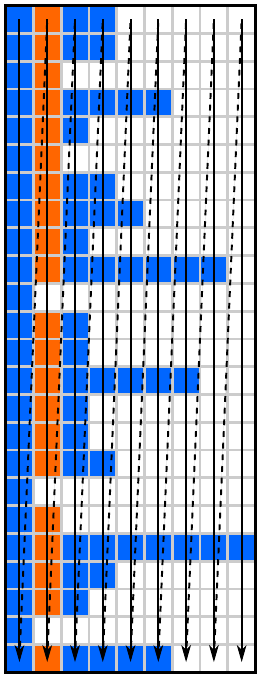}}
\hfill
\subfloat[JDS]{\label{fig:jds}\includegraphics*[height=0.25\textheight]{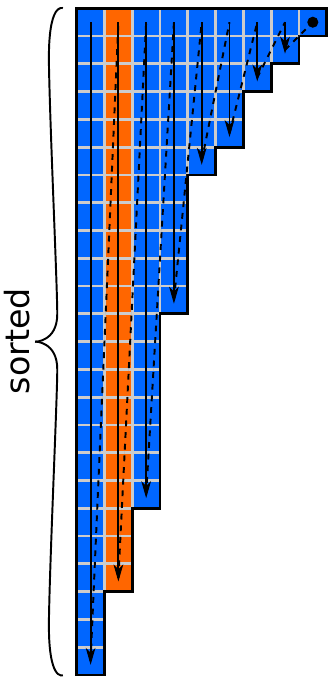}}
\caption{Derivation of standard sparse matrix storage formats. 
  Arrows indicate the storage order of matrix values
  and column indices. The highlighted non-zeros form a column of
  entries in the storage formats; in case of CRS, these are not
  stored consecutively.}
\label{fig:oldformats}
\end{figure}

In \autoref{fig:oldformats} we sketch the most popular sparse matrix storage
formats on CPUs (\autoref{fig:crs}), GPGPUs (\autoref{fig:ellpack}) 
and vector computers (\autoref{fig:jds}). There are strong differences 
between these formats in terms of the 
storage order of the non-zero entries, the use of padding (ELLPACK),
and the row reordering (JDS), due to the peculiarities of each
hardware platform. These differences make it tedious to 
use heterogeneous systems efficiently. In the
following we identify a unified low-overhead storage format, 
which is designed to be efficient on the three classes of
compute devices considered in this work. 
Guided by the equivalence of SIMD and SIMT (warp)
execution, we analyze the overhead and benefit of SIMD
vectorization strategies for the CRS format and for an improved variant of
the ELLPACK scheme.

\subsection{Compressed Row Storage (CRS)}
The CRS data format is a cache-friendly layout ensuring consecutive
data access to the matrix elements and the column indices. The
C version of a CRS spMVM kernel is given in
\autoref{listing:crs}.
\lstset{caption={CRS spMVM kernel}\label{listing:crs}}
\begin{lstlisting}[float=tbp]
for(i = 0; i < N; ++i) {
  for(j = rpt[i]; j < rpt[i+1]; ++j) {
      y[i] += val[j] * x[col[j]];
  }
}
\end{lstlisting}

The non-zero matrix entries are stored row by row in the array
\verb.val[]., and their original column indices are put 
in \verb.col[].. The starting offsets of all rows are available 
in the array \verb.rpt[]..  A sketch of the CRS 
storage scheme for the matrix in
\autoref{fig:spm} is shown in \autoref{fig:crs}.


Efficient SIMD vectorization requires consecutive data access for
optimal performance. Thus the inner loop in \autoref{listing:crs},
which runs over all non-zero entries of each row, is the target for
vectorization.  Applying four-way ``modulo unrolling'' to this loop, we can formulate
the CRS spMVM kernel in a SIMD-friendly way, tailored for the AVX
register width of four elements as used by the Intel SNB processor (see
\autoref{listing:crs_vec}).

\noindent\begin{minipage}[b]{.47\textwidth}
\centering
\lstset{caption={CRS spMVM kernel with four-way modulo unrolling.}\label{listing:crs_vec}}
\begin{lstlisting}
for(i = 0; i < N; ++i) 
{
  tmp0 = tmp1 = tmp2 = tmp3 = 0.; 
  for(j = rpt[i]; j < rpt[i+1]; j+=4)
  {
    tmp0 += val[j+0] * x[col[j+0]];
    tmp1 += val[j+1] * x[col[j+1]];
    tmp2 += val[j+2] * x[col[j+2]];
    tmp3 += val[j+3] * x[col[j+3]];
  }
  y[i] += tmp0+tmp1+tmp2+tmp3;%\label{l:hadd}%
  // remainder loop
  for(j = j-4; j < rpt[i+1]; j++)%\label{l:sr_start}%
    y[i] += val[j] * x[col[j]];%\label{l:sr_end}%
}
\end{lstlisting}
\end{minipage}
\hfill
\begin{minipage}[b]{.47\textwidth}
\lstset{caption={SELL-4-$\sigma$ spMVM kernel with four-way unrolling.\label{listing:sell}}}
\begin{lstlisting}
for(i = 0; i < N/4; ++i) 
{

  for(j = 0; j < cl[i]; ++j) 
  {
    y[i*4+0] += val[cs[i]+j*4+0] * 
              x[col[cs[i]+j*4+0]];
    y[i*4+1] += val[cs[i]+j*4+1] * 
              x[col[cs[i]+j*4+1]];
    y[i*4+2] += val[cs[i]+j*4+2] * 
              x[col[cs[i]+j*4+2]];
    y[i*4+3] += val[cs[i]+j*4+3] *
              x[col[cs[i]+j*4+3]];
  }
}
\end{lstlisting}
\end{minipage}

The compiler can often do this by itself and
vectorize the bulk loop such that the body is
executed in a SIMD-parallel manner, e.g., \verb!tmp[0,...,3]! is
assigned to a single AVX register and \verb!val[j+0,...,j+3]! is
loaded with a single instruction to another register. The initial loop
peeling to satisfy alignment constraints is omitted for brevity.

The same strategy is chosen by the Intel compiler for the
vectorization of the basic CRS code on the Intel Phi with an
appropriate choice of unrolling factor (eight instead of four; see
\cite{saule13}).

\subsection{Analysis of the CRS format}
\label{sec:crsprobs}

Vectorized execution of the CRS spMVM may be inefficient, especially
for matrices with few non-zeros per row ($N_\mathrm{nzr}$).
$N_\mathrm{nzr}$-independent overheads of the
vectorized code, like the ``horizontal'' add operation (line
\ref{l:hadd} in \autoref{listing:crs_vec}) or the scalar remainder
loop (lines \ref{l:sr_start}--\ref{l:sr_end}) may then eat up the
performance gained by vectorization.  Note that these particular
costs grow with the SIMD width.

Especially on the Intel Phi, handling
of scalar overheads like a remainder loop may be expensive: Even though 
almost all SIMD operations can be masked to
``emulate'' scalar execution, there is an additional penalty for setting
up the mask and executing a separate instance of the loop body.
The worst case occurs when the row length is in the order of (or even
smaller than) the SIMD width.  In this case, the amount of
non-vectorized and/or inefficiently pipelined work introduces a
significant overhead. For instance, on the Intel Phi a total of 16
single-precision (or eight double-precision) values can be processed
with a 512-\BIT\ SIMD instruction at a time. A good utilization of the
SIMD lanes thus demands an even larger $N_\mathrm{nzr}$ than on the
Intel SNB.  Additionally, alignment constraints may require some loop
peeling, which further reduces the SIMD-vectorized fraction. In
summary, on a wide-SIMD architecture the average number of non-zeros per
row needs to be substantially larger than the SIMD width of the
compute device.

The rather large SIMD width of the GPGPU architecture together with the
cost of reduction overhead even within a warp (see also the discussion in
\autoref{sec:matrices}) immediately rules out the CRS format if
SIMD/SIMT execution is performed along the inner loop. Parallelizing the
outer loop eliminates this problem but destroys load coalescing, since
threads within a warp operate on different rows and access elements 
concurrently that are not consecutive in memory. Hence, CRS is a
bad choice on GPGPUs in any case \cite{bellgarland09}.


\subsection{Sliced ELLPACK and \SELLCS}\label{sect:sell}

\begin{figure}[tb]
\subfloat[SELL-6-1, $\beta=0.51$][SELL-6-1,\\$\beta=0.51$]{\label{fig:sell}\includegraphics*[height=0.25\textheight]{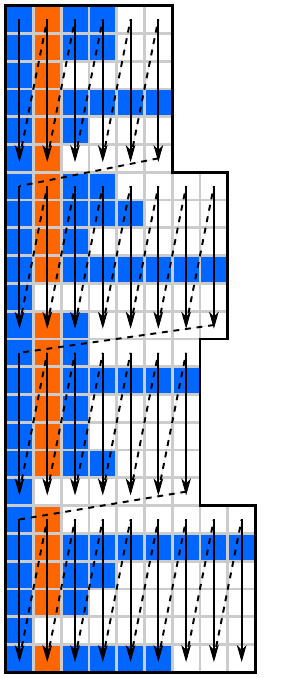}}
\hfill
\subfloat[SELL-6-12, $\beta=0.66$][SELL-6-12,\\$\beta=0.66$]{\label{fig:sell12}\includegraphics*[height=0.25\textheight]{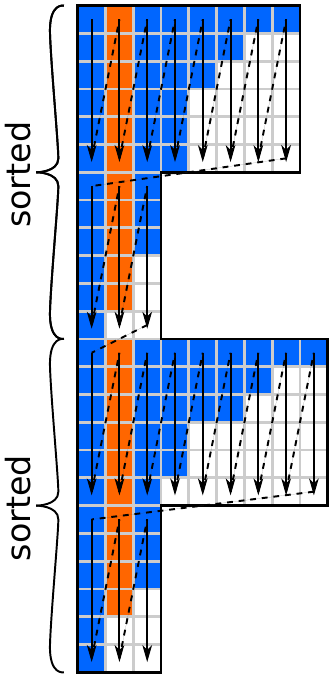}}
\hfill
\subfloat[SELL-6-24, $\beta=0.84$][SELL-6-24,\\$\beta=0.84$]{\label{fig:sell24}\includegraphics*[height=0.25\textheight]{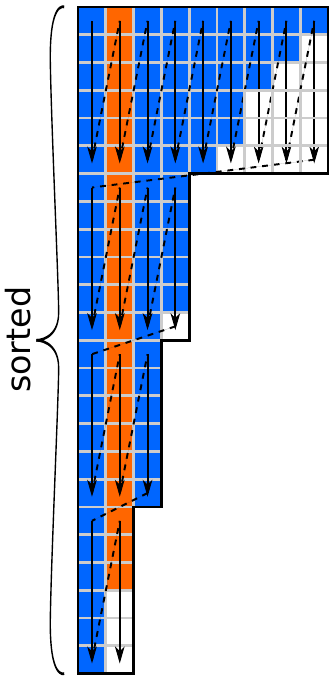}}
\hfill
\begin{minipage}[b]{.3\textwidth}
\caption{Variants of the \SELLCS\ storage format for the matrix 
 structure in \autoref{fig:spm}. Arrows indicate the storage
 order of matrix values and column indices.\label{fig:newformats}\vspace*{-0.6cm}}
\end{minipage}
\end{figure}
The ELLPACK format addresses the problems of CRS for GPGPUs.
A Column-wise data layout and padding all rows to the same
length (see \autoref{fig:ellpack}) allows
row-wise thread parallelization and coalesced memory access to the
matrix data. Of course, the number of rows must also be padded to
a multiple of the warp size (not shown in \autoref{fig:ellpack}). 
Since ELLPACK may incur substantial overhead for padding
(see white boxes in \autoref{fig:ellpack}), Monakov et al.\
\cite{monakov10} have proposed a variant called ``Sliced ELLPACK,''
which substantially reduces this overhead and increases data locality
between successive column computations within a warp.

The main idea is to cut the ELLPACK data layout into equally-sized
``chunks'' of rows with $C$ rows per chunk. Rows are zero-padded to
match the length of the longest row \emph{within their chunk},
reducing the padding overhead substantially as compared to ELLPACK.
Then all elements within a padded chunk are stored consecutively in
column-major order, and all chunks are consecutive in memory (see
\autoref{fig:sell}). Unless all rows in the chunk
are of equal length, there may still be a substantial penalty in terms
of data storage, which will be discussed in
\autoref{sec:sellprobs}. In addition, the number of matrix rows $N$
must be padded to a multiple of $C$.
We call this format ``\SELLCS,'' since it is parametrized
by the chunk size $C$ and a ``sorting scope'' $\sigma$, which
will be explained in \autoref{sect:sellpit}. For the remainder of this section we
assume no sorting ($\sigma=1$).

\autoref{listing:sell} shows the C version of the spMVM for SELL-4-$\sigma$, 
unrolled to match the SIMD width of an AVX-capable
processor. For reference, a version of the SELL-4-$\sigma$ spMVM with AVX intrinsics 
is shown in \autoref{listing:sell_avxintr}.
The matrix entries and their column indices are stored in
arrays \verb.val[]. and \verb.col[].. In addition, the starting offset
of each chunk is stored in \verb.cs[]. and the width of each chunk,
i.e., the length of the longest row in the chunk, is stored in
\verb.cl[]. (\verb.cl[i]. is equal to (\verb.cs[i+1].-\verb.cs[i].)/$C$).

In contrast to the CRS spMVM, the \SELLCS\ kernel has a vectorizable
inner loop without a reduction operation, so the ``horizontal add'' is
not required. Also the remainder loop handling is obsolete since $N$
is padded to a multiple of $C$. A direct comparison with CRS shows
that the \emph{inner} loop unrolling in the \SELLCS\ kernel
corresponds to \emph{outer} loop unrolling in CRS but ensures cache
locality and eases alignment and coalescing constraints, since the
matrix data accessed in the inner loop iterations is consecutive in
memory.  The \SELLCS\ kernel can be vectorized by the compiler or
through the use of C intrinsics on Intel systems. Thus, \SELLCS\ is a
promising candidate for delivering high efficiency on a variety of
compute devices.

The optimal choice of $C$ needs to take into account both the padding 
overhead of the \SELLCS\ format and hardware-specific restrictions.
SELL-$N$-1 is identical to ELLPACK (\verb.cl[]. and \verb.cs[]. are
not strictly needed in this case), and has maximum padding overhead
as discussed, e.g., in \cite{spmvm12-KHW}. The other extreme case SELL-1-1 is
equivalent to CRS, and there is no padding at all.
Hence, it is crucial to choose $C$ as small as possible but still
compatible with architectural requirements. On the \nvidia\ K20, the
reasonable (minimum) choice is $C=32$, i.e., each chunk is executed by one warp.
According to the equivalence of SIMD and SIMT execution, a
first choice for $C$ on CPUs would be the SIMD register width in units of
the matrix value data size  (e.g., $C=4$ as shown in \autoref{listing:sell} 
for an AVX-capable processor in double precision in plain C and in 
\autoref{listing:sell_avxintr} with compiler intrinsics). On the Intel Phi one would
naively set $C=8$; however, all vectorized data accesses need to be
512-\BIT\ aligned. This leads to the hardware-specific constraint that
\bq
C \cdot \mbox{\texttt{cl[i]}} \cdot \mathrm{min(\mathtt{sizeof(*val)},\mathtt{sizeof(*col)})}
\eq
has to be a multiple of 64 \BYTES\ on the Intel Phi, where \verb.cl[i]. 
is the length of the $i$-th chunk.
In our case (double-precision matrix, four-byte integer index) 
this condition is fulfilled with 
\bq
C = 64/\min{(8,4)} = 16
\eq
independently of \verb.cl[i]., so we choose $C=16$ for Intel Phi.

Note that on a heterogeneous system, different minimal values of $C_i$
may apply to each component architecture $A_i$. An obvious solution to
this issue in order to obtain a consistent format is to choose the
global chunk height $C = \max_i(C_i)$. See \autoref{sect:UDL} for a
discussion of a unified data format for all architectures.


\subsection{Analysis of the \SELLCS\ format}\label{sect:sellpit}
\label{sec:sellprobs}
In order to quantify the overhead incurred by the zero-padding in the
\SELLCS\ format we define the ``\cocc'' $\beta$.
It is the fraction of ``useful'' matrix data entries, i.e.,
the ratio between the number of non-zero matrix
elements $N_{\mathrm{nz}}$ and the elements stored in the \SELLCS\ format:
\bq\label{beta}
\begin{split}
\beta=\frac{N_{\mathrm{nz}}}{\sum_{i=0}^{N_\mathrm c} C \cdot \mbox{\texttt{cl[i]}}}\eos \\
\end{split}
\eq 
Here, $N_\mathrm c$ is the number of chunks for the matrix, 
\bq\label{nc}
N = N_\mathrm c \cdot C\cma 
\eq 
and \verb.cl[i]. is defined as above:
\bq\label{max} \mbox{\texttt{cl[i]}}=\max\limits_{k=iC}^{(i+1)C-1} \text{rowLen[$k$]}
\eq 
The $\beta$ values for all test matrices can be found in
\autoref{table:matrices}. The meaning of $\sigma$ will be
explained below.

The minimal value for $\beta$ (worst case scenario) indicates a matrix
structure for which the \SELLCS\ data transfer overhead is at a maximum. Such
a matrix has a single (fully populated) row with $N$ non-zeros in each
chunk, and only a single non-zero in all other rows of the same chunk.  
In this case, $C\times N$ elements have to be loaded per chunk
while only $N+C-1$ elements are actually non-zero:
\bq\label{betaworst}
\begin{split}
\beta_\mathrm{worst}&=\frac{\sum_{k=0}^{N_\mathrm c}(N+C-1)}{\sum_{k=0}^{N_\mathrm c} CN} \\
	&=\frac{N+C-1}{CN} \stackrel{N\gg C}{\longrightarrow}\frac 1C \eos
\end{split}
\eq 

In contrast to this, a constant row length within each chunk (the row
length does not have to be constant globally) leads to the best case
scenario with $\beta=1$, since no zero-padding elements have to be 
transferred.

A small $\beta$ can be increased by sorting the matrix rows by
row length in descending order, so that rows of equal length end up close to
each other. Obviously, the overhead becomes minimal when sorting the
matrix rows globally, as shown in \autoref{fig:sell24}. In this case, $\beta \approx 1$ and the matrix
format is identical to pJDS \cite{spmvm12-KHW}, which can be
considered as a zero-padded version of the JDS format (see
\autoref{fig:jds}) with appropriate $C$.  However, when sorting matrix
rows globally there is a chance that the access pattern to the RHS
vector changes substantially and spatial or temporal
locality arising from the physical problem is destroyed. This may lead
to an increase in code balance (more data transfers are needed per
flop) and, as the kernel performance is already limited by data
transfers, to a performance drop.  See \autoref{sect:perfmodels} below
for a quantification of such effects using suitable performance
models.

A way to ameliorate this problem is to not sort the matrix rows
globally but only within chunks of $\sigma$ consecutive
rows. Typically, this ``sorting scope'' $\sigma$ is chosen to be a
multiple of $C$; if $\sigma$ is a divisor of $C$, there is no effect
on $\beta$. Here we restrict our analysis to powers of two for $C$ and
$\sigma$ (it is certainly not ruled out that a specific,
  non-power of two choice of $\sigma$ might be advantageous for a
  specific matrix).  The effect of local sorting is shown in
\autoref{fig:sell12} for $C=6$ and $\sigma=12$. The ``optimal''
$\sigma$, i.e., for which the RHS access is still
``good'' but which leads to a sufficiently large $\beta$, is usually
not known a priori. Only for very regular matrices can this problem be
solved exactly: For the worst-case matrix with $\beta$ as given in
\autoref{betaworst}, $\sigma=C^2$ results in a perfect $\beta = 1$. In
this case, there is one chunk with length $N$ and $C-1$ chunks of
length one within the scope of $\sigma=C^2$ rows. 

At this point it has to be noted that when sorting the matrix rows,
the column indices need to be permuted accordingly in most of the
application cases. This has two major reasons: First, in iterative
solvers the algorithm usually switches after each iteration between the
``input'' and ``output'' vectors of the previous spMVM operation. Thus
these schemes often work in the permuted indices space.  Second,
possible ``matrix bandwidth escalation'' of the non-zero pattern due
to row re-ordering may be averted by permuting the column indices. The
matrix bandwidth is the maximum distance of non-zero entries from the
main diagonal.

Sorting the matrix rows is part of the preprocessing step and has to
be done only once.  Assuming that a large number of spMVM operations
will be executed with the sorted matrix, the relative overhead of the
sorting itself can usually be neglected.  Furthermore, since we only
sort inside a certain limited scope, the cost of sorting a single
scope is small and parallelization across different scopes is
straightforward.

On GPGPUs, the \SELLCS\ format enables a specific optimization: Due to
the fact that there is one dedicated thread per matrix row, it is easy
to avoid loading zero matrix elements by letting each thread run only
until the actual row length is reached.  This makes the data format
equivalent to the ``Sliced ELLR-T'' format as introduced by Dziekonski
et al.\ \cite{dziekonski11} (with the number of threads running per
row set to $T=1$).  However, there is still a penalty for low-$\beta$
matrices on GPGPUs as the resources occupied by the threads of a warp
are only available after the longest-running thread of this warp has
finished.

\subsection{General performance issues of spMVM}
\label{sect:gperfiss}

Beyond the issues of vectorization and excess data transfers, 
indirect access to the RHS vector \verb.x. may further impede
the performance of the spMVM kernel for reasons unconnected
to a specific data storage format.

First, performance will drop significantly if the elements of
\verb.x. accessed in consecutive inner spMVM loop iterations are not
close enough to each other, so that inner cache levels or ``load
coalescing'' cannot be used and main memory data access becomes
irregular. A rather general approach to address this problem is to
reduce the matrix bandwidth by applying a bandwidth reduction
algorithm, such as ``Reverse Cuthill McKee'' (RCM)
\cite{Cuthill:1969:RBS:800195.805928}. Such transformations are
outside the scope of this work, but the impact of non-consecutive
accesses can be explored in more detail using performance 
models (see \autoref{sect:perfmodels}).

Second, moving the elements of the RHS vector \verb.x.  into a SIMD
register may come along with a large instruction overhead. Up to now,
x86-based CPU instruction sets (including AVX) do not provide a
``gather'' instruction, and loading the elements of \verb.x. has to
be done with scalar loads. Thus, filling an AVX register with four
elements of \verb.x. requires at least five instructions in total (one
to load the four consecutively stored indices and four to fill the
vector registers with the values of \verb.x.; in practice, the number
is even larger since the individual slots of a SIMD register can
usually not be freely addressed).  However, Intel's MIC architecture
does provide a ``gather'' instruction. It can fetch multiple data
items residing in the same cache line from memory to a vector register
even if the addresses are not consecutive. This potentially enhances
performance for loading the elements of \verb.x. compared to scalar
loads. However, the actual benefit depends on the locality of the
matrix entries in a single row. In the worst case, i.e., if all
gathered items reside in different cache lines, one gather instruction
per load is required and the whole operation is basically scalar
again. Note that the adverse effects of instruction overhead will only be
visible if no other resource such as main memory bandwidth (the most
promising candidate for spMVM) already limits the attainable
performance. This is true on any architecture.

\section{Performance models}\label{sect:perfmodels}

For large data sets, the spMVM is strongly memory-bound. The spMVM
kernels in Listings \ref{listing:crs}, \ref{listing:crs_vec}, and
\ref{listing:sell} are characterized mainly by data streaming (arrays
\verb.val[]. and \verb.col[].) with partially indirect access (RHS
vector \verb.x.). Assuming no latency effects and infinitely fast
caches, it is possible to establish roof\/line-type performance models
\cite{Williams:2009:RIV:1498765.1498785}.

The code balance, i.e., the number of bytes transferred over the
memory interface per floating-point operation, can be deduced from
\autoref{listing:crs} for \emph{square matrices}: 
\cite{spmvm12-KHW,lspp11,6267822} 
\bq
B_\mathrm{CRS}^\mathrm{DP}=\left(\frac{v_\mathrm{mat}+v_\mathrm{RHS}+v_\mathrm{LHS}}{2\,\FLOPS}\right)\cma
\eq
where $v_\mathrm{mat}$ accounts for reading the matrix entries and
column indices, $v_\mathrm{RHS}$ is the traffic incurred by reading
the RHS vector (including excess traffic due to insufficient spatial
and/or temporal locality), and $v_\mathrm{LHS}$ is the data volume for
updating one LHS element. Assuming double precision matrix and vector
data and four-byte integer indices we have
$v_\mathrm{mat}=(8+4)\,\BYTES$ and $v_\mathrm{LHS}=16\,\BYTES/N_\mathrm{nzr}$.  The
efficiency of the RHS data access is quantified by the parameter
$\alpha$ in $v_\mathrm{RHS}=8\alpha\,\BYTES$. Hence, we get:
\bq\label{crsbalance}
\begin{split}
B_\mathrm{CRS}^\mathrm{DP}&=\left(\frac{8+4+8\alpha+16/N_\mathrm{nzr}}{2}\right)\frac{\BYTES}{\FLOP} \\
&=\left(6+4\alpha+\frac{8}{N_{\mathrm{nzr}}}\right)\frac{\BYTES}{\FLOP}\eos
\end{split}
\eq 
The value of $\alpha$ is governed by a subtle interplay between the
matrix structure and the memory hierarchy on the compute device: If
there is no cache, i.e., if each load to the RHS vector goes to
memory, we have $\alpha=1$. A cache may reduce the balance by some
amount, to get $\alpha<1$.  In the ideal situation when
$\alpha=1/N_\mathrm{nzc}$ (with $N_\mathrm{nzc}$ the average number of non-zero elements per column
and $N_\mathrm{nzc} = N_\mathrm{nzr}$ for square matrices), 
each RHS element has to be loaded only once
from main memory per spMVM\footnote{This corresponds to the $\kappa=0$
  case in \cite{lspp11}.}. The worst possible scenario occurs when the
cache is organized in cache lines of length $L_\mathrm C$ elements,
and each access to the RHS causes a cache miss. In this case we have
$\alpha=L_\mathrm C$, with $L_\mathrm C=8$ or $16$ on current
processors. As already discussed in \autoref{sect:gperfiss}, the
locality of the RHS vector access and, consequently, the value of
$\alpha$ can be improved by applying matrix bandwidth reduction
algorithms. Note also that, depending on the algorithm and the problem
size, the RHS vector may reside in cache for multiple subsequent spMVM
kernel invocations, although the matrix must still be fetched from
memory. In this special case we have $\alpha=0$.

The CRS-based roof\/line model \autoref{crsbalance} must be modified for the
\SELLCS\ data format.  As discussed in \autoref{sec:sellprobs},
additional data is loaded and processed if the row lengths vary inside
a chunk.
The reciprocal of the \cocc\ $\beta$ quantifies
the format-inherent average data traffic per
non-zero matrix element. Note the excess traffic for $\beta<1$ only
arises for the matrix value and column index and not for the RHS
element.  This is because all padded column indices are set to zero;
thus, only the 0-th RHS element is accessed for all padded elements
and the corresponding relatively high access frequency will ensure
that this element stays in cache. The code balance for \SELLCS\ is then
\bq\label{sellbalance}
\begin{split}
B_\mathrm{SELL}^\mathrm{DP}(\alpha,\beta,N_\mathrm{nzr})&=\left(\frac{1}{\beta}\left(\frac{8+4}{2}\right)+
  \frac{8\alpha+16/N_\mathrm{nzr}}{2}\right)\frac{\BYTES}{\FLOP} \\
  &=\left(\frac{6}{\beta}+4\alpha+\frac{8}{N_\mathrm{nzr}}\right)
     \frac{\BYTES}{\FLOP}\eos
\end{split}
\eq 
The roof\/line model can now be used to predict the maximum achievable spMVM
performance:
\bq\label{roof_general} 
P(\alpha,\beta,N_\mathrm{nzr},b)=
\frac{b}{B_\mathrm{SELL}^\mathrm{DP}(\alpha,\beta,N_\mathrm{nzr})}\eos
\eq
Here, $b$ is the achievable memory bandwidth as determined by a
suitable microbenchmark, e.g., one of the two benchmarks discussed in
\autoref{sec:hardware}. Using $\beta=1$ in \autoref{roof_general} we
obtain the analogous expression for CRS format.

As a special case we focus on the $\alpha=1/N_\mathrm{nzr}$ scenario, which
has been described above. On the Intel SNB processor, whose LLC of 20
\MiB\ can (in theory) hold at least a single vector of all matrix
sizes in \autoref{table:matrices}, this is usually a valid
assumption. Then the performance model reads:
\bq\label{roof_alpha} 
P(1/N_\mathrm{nzc},\beta,N_\mathrm{nzr},b)=
\frac{b}{\left(6/\beta+4/N_\mathrm{nzc}+8/N_\mathrm{nzr}\right)\,\footnotesize\frac{\BYTES}{\FLOP}}\eos
\eq

For square matrices with a sufficiently large number of non-zeros per
row ($N_\mathrm{nzr} \gg 12 $) one finally arrives at the best
attainable performance level for spMVM operations considered in this
work: 
\bq\label{best_veccache}
\bar{P}=\frac{b\beta}{6\,\footnotesize\frac{\BYTES}{\FLOP}}\eos
\eq 
Note that these estimates are based on some optimistic
assumptions which may sometimes not hold in reality: Main memory bandwidth is
the only performance limiting factor, data access in cache is
infinitely fast, the cache replacement strategy is optimal, and no latency
effects occur. Nevertheless, \autoref{best_veccache} provides an upper
bound for spMVM performance on all compute devices if the matrix data
comes from main memory.



In general, when RHS accesses cannot be neglected, the code balance
depends on $\alpha$, which can only be predicted in very simple
cases. However, $\alpha$ can be determined by measuring the used
memory bandwidth or data volume of the spMVM kernel and setting the
code balance equal to the ratio between the measured transferred
data volume $V_{\mathrm{meas}}$ and the number of executed ``useful''
flops, $2 \times N_\mathrm{nz}$.  Note that this is only possible if
the code is limited by memory bandwidth. For \SELLCS\ we then obtain
\bq
B_\mathrm{SELL}^\mathrm{DP} = \left(\frac{6}{\beta}+4\alpha+
  \frac{8}{N_\mathrm{nzr}}\right)\frac{\BYTES}{\FLOP} = 
  \frac{V_{\mathrm{meas}}}{N_{\mathrm{nz}}\cdot 2\,\FLOPS}\cma
\eq
which can be solved for $\alpha$:
\bq\label{eq:sellalpha}
\alpha
= 
  \frac{1}{4}\left(\frac{V_{\mathrm{meas}}}{N_{\mathrm{nz}}\cdot 2\,\BYTES}
  -\frac{6}{\beta}-\frac{8}{N_{\mathrm{nzr}}}\right) \eos
\eq 
The corresponding CRS values can again be retrieved by setting $\beta=1$.

\section{Performance results and analysis}\label{sec:perfresults}

In our experiments all matrices and vectors are of \verb.double.
type and all indices are four-byte integers. The execution time
of a series of spMVM operations has been measured to account for
possible caching effects and to reduce the impact of finite timer
accuracy. The performance analysis always uses a full compute device
(one chip).

The OpenMP scheduling for the Intel architectures has been set following
a simple heuristic based on the matrix memory footprint and its
coefficient of variation ($\zeta$, standard deviation divided by mean) regarding row lengths:
\bq\label{eq:cv}
\zeta = \frac{\sqrt{\frac{\sum_{i=0}^{N}(\mathrm{rowLen[i]}-N_\mathrm{nzr})^2}{N}}}{N_\mathrm{nzr}}
\eq
If the matrix fits in the LLC or $\zeta < 0.4$ we use 
``\verb.STATIC.'' scheduling, else we use ``\verb.GUIDED,1.'' scheduling.
Especially on the Intel Phi the OpenMP scheduling
may have significant impact on the performance and needs to be chosen
with care.

The clock frequency of the Intel SNB was fixed to 2.7\,\GHZ, and $C_\mathrm{arch}=4$
has been chosen in accordance with the AVX register width. 
Eight OpenMP threads were used and
Simultaneous Multi-Threading (SMT) was disabled.

For the Intel Phi a chunk height of $C_\mathrm{arch}=16$ has been selected to ensure
both SIMD vectorization and alignment constraints following the
discussion in \autoref{sect:sell}. Best performance was generally
achieved on all 60 cores with three threads per core (Saule et al.
\cite{saule13} came to the same conclusion). 
The large unrolling factor of at least $C=16$ for the \SELLCS\ kernel 
makes the loop body rather bulky and hard to efficiently vectorize by the
compiler. Thus, the \SELLCS\ kernel for Intel Phi has been implemented
using MIC compiler intrinsics as shown in \autoref{listing:sell_micintr},
ensuring efficient vectorization.

On the \nvidia\ K20, $C=32$ was set according to the hardware-specific
warp size for optimal load coalescing and data alignment. For
execution of the CUDA code, a CUDA block size of $256$ has been used
unless otherwise noted. A single thread was assigned to each row.

\subsection{Unified data layout performance}
\label{sect:UDL}
\begin{figure}[tbp]
\centering
\subfloat[Memory-bound]{\label{fig:perf_mem}\includegraphics[width=\textwidth]{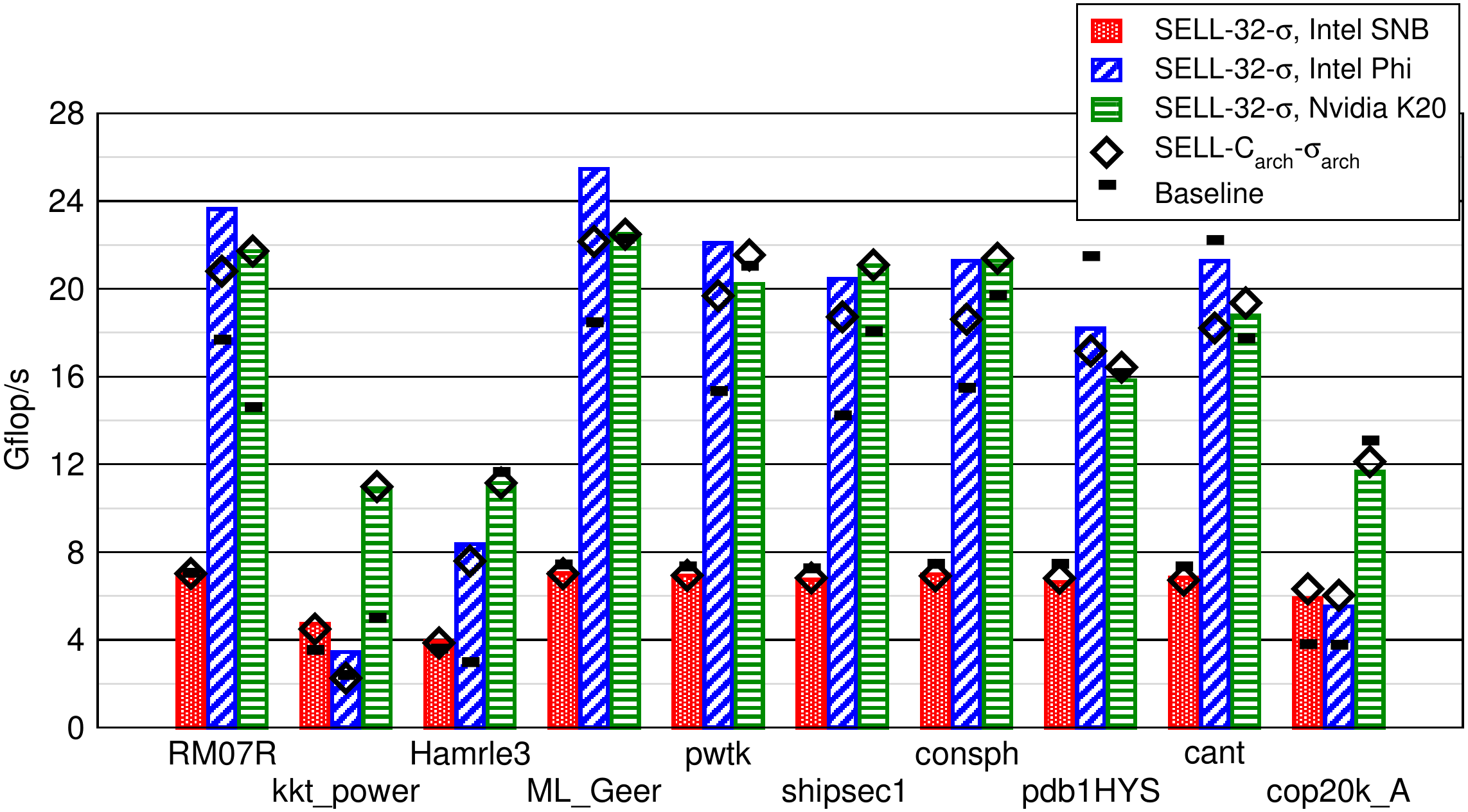}}\\
\subfloat[Cache-bound]{\label{fig:perf_cache}\includegraphics[height=16.9em]{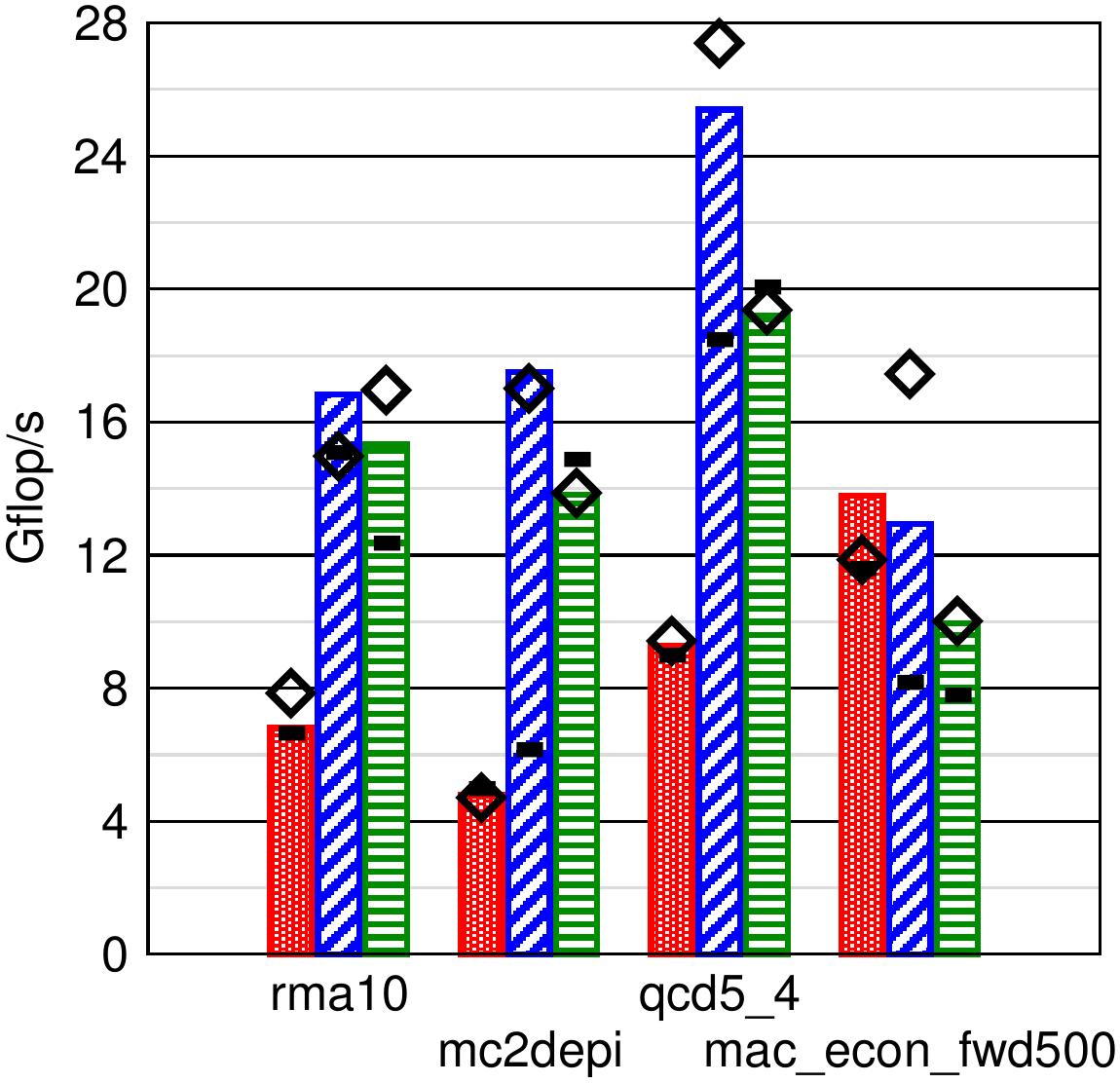}}
\hfill
\subfloat[Pathological]{\label{fig:perf_spec}\includegraphics[height=16.9em]{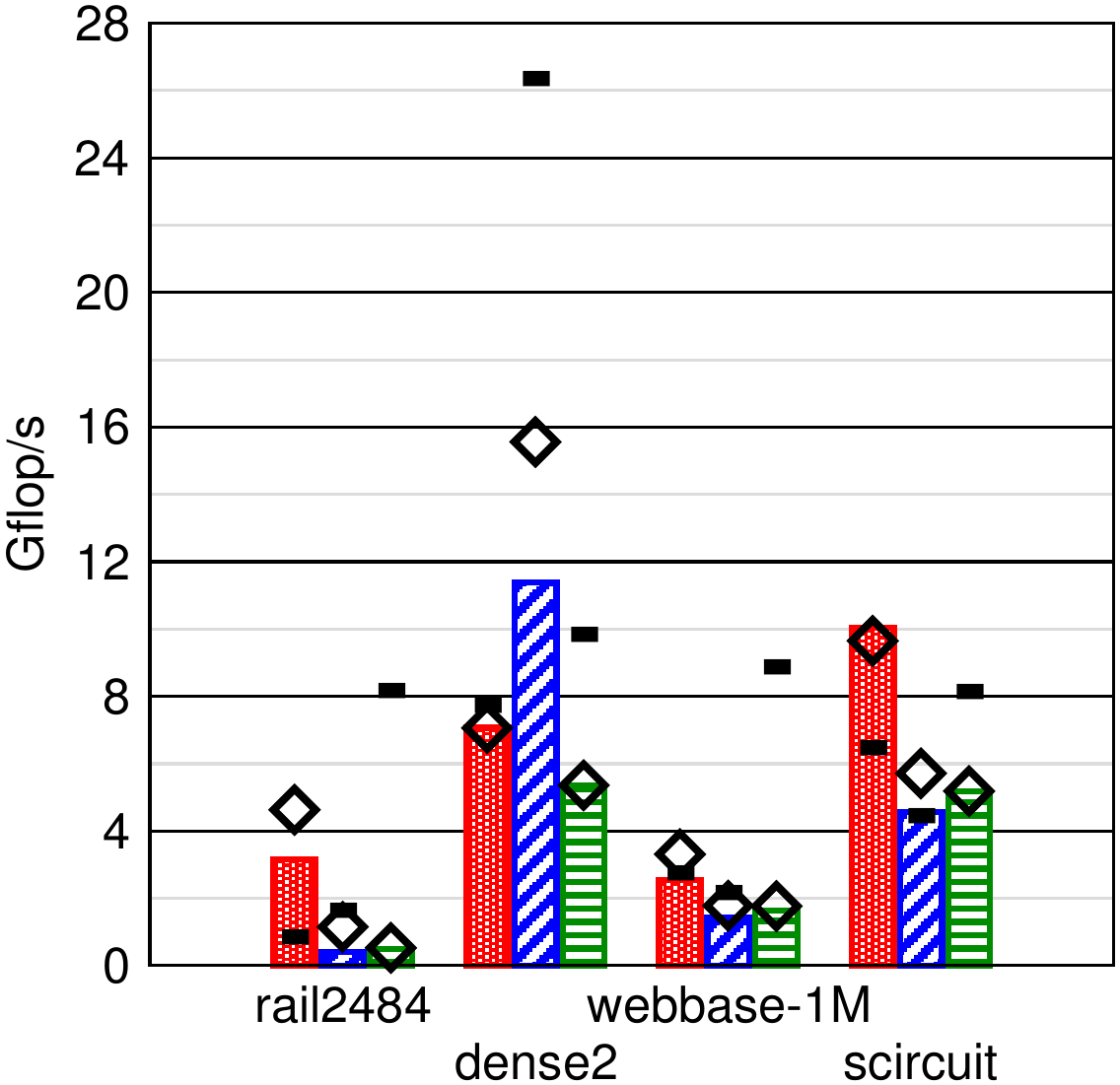}}
\caption{Overview of the spMVM performance for all test matrices.
	CRS is compared to the \SELLCS\ format using different sorting
  scopes $\sigma$ for the latter. 
  For hardware-specific chunk sizes $C$ see text.\\
  (a) Matrices with
  $N_\mathrm{nz}>2.5 \times 10^6$ (memory bound on all devices); the four
  matrices on the left constitute ``corner cases''.\\ 
  (b) Matrices with $N_\mathrm{nz}<2.5 \times 10^6$ (fit into the LLC of Intel Phi at least).\\
  (c) Matrices with specific characteristics (see text for discussion).}
\label{fig:performance_all}
\end{figure}

The performance of the various data layouts across all hardware
platforms was investigated by classifying the test matrices into three
groups. \autoref{fig:perf_mem} shows a survey of the
matrices for which the complete memory footprint of the spMVM data
($\gtrsim 12\times N_\mathrm{nz}$ \BYTES) is larger than any LLC on
all compute devices (i.e., $N_\mathrm{nz} > 2.5\times 10^6$). In
contrast to these strongly memory bound cases, we present in
\autoref{fig:perf_cache} all matrices which can in theory run
completely out of the LLC at least on the Intel Phi,
which has the largest LLC of all.
The last group as shown in \autoref{fig:perf_spec} is constituted by 
matrices with specific characteristics where \SELLCS\
reveals its shortcomings.

The baseline performance has been obtained with the vendor-supplied sparse
linear algebra libraries using the ``standard'' data format. Concretely, we used Intel MKL 
11.0 with CRS for Intel Sandy Bridge/Xeon Phi (COO for the non-square matrix rail4284 as MKL's CRS
cannot handle non-square matrices) and \nvidia\ cuSPARSE 5.0 with HYB (cf. \cite{bellgarland09}) on the \nvidia\ K20.

In all groups, the ``SELL-32-$\sigma$'' bar shows the performance of the unified data layout for each
architecture.
The sorting scope has been chosen to the value in the range ($1\leq \sigma \leq 2^{17}$) where the
average relative performance with respect to the baseline over all three architectures has its maximum.

``SELL-$C_{\mathrm{arch}}$-$\sigma_{\mathrm{arch}}$'' results show the
performance obtained with the hardware-specific $C_{\mathrm{arch}}$ (as described above) and the 
optimal $\sigma$ in the same range as described above for each matrix and architecture.


The four corner case matrices (cf. \autoref{sec:matrices}) have been
chosen such that we can test the \SELLCS\ scheme for the limits of
small/large number of non-zeros per row ($N_\mathrm{nzr}$) and
high/low \cocc\ ($\beta$) of the original (unsorted) matrix (see
\autoref{table:matrices} for exact values).  They are shown on the far
left in \autoref{fig:perf_mem}.
On Intel SNB the ``native'' CRS layout no longer has any advantage compared to the
SIMD-vectorized \SELLCS\ format.  On the Intel Phi, \SELLCS\ is far superior to
CRS for all matrices. The advantage becomes most pronounced for the
``low-$N_\mathrm{nzr}$'' matrix Hamrle3, where CRS is extremely slow
on this wide-SIMD architecture due to the problems discussed in
\autoref{sec:crsprobs}.
\SELLCS\ outperforms HYB in cases where $\beta_{\sigma=1}$ is rather low.
There are two possible reasons for that, depending on the threshold chosen by cuSPARSE for
automatic partitioning of the HYB matrix: Either (for a low threshold) the overhead 
induced by the irregular (COO) part of the HYB matrix gets too large or (for a high threshold) the overhead
from zero fill-in in the regular (ELL) part of the HYB matrix gets too large.

In general, \SELLCS\ with optimal sorting attains best performance
on all architectures, with highest impact (as compared to no sorting)
for the matrices with worst \cocc, e.g RM07R and
kkt\_power. 

In case of the larger set of memory-bound test matrices from the ``Williams
group'' (right part of \autoref{fig:perf_mem}) the \SELLCS\
format also provides best performance for all matrices and
architectures if an optimal sorting scope is used. 
For data
sets that completely fit into the LLC of Intel platforms (\autoref{fig:perf_cache}), the \SELLCS\ format
substantially outperforms CRS. The much lower instruction
overhead of vectorized \SELLCS\ vs. (vectorized) CRS boosts performance
on the Intel Phi by $1.5${}$\times$ to $4${}$\times$ for the matrices shown in
\autoref{fig:perf_cache}. 
On the Intel SNB, \SELLCS\ shows similar benefits
for the matrix mac\_econ\_fwd500 which can be held in its LLC.
  

\subsection{Detailed performance analysis} \label{sect:det_perf_analysis}

\begin{figure}[tbp]
\centering
\includegraphics[width=\textwidth]{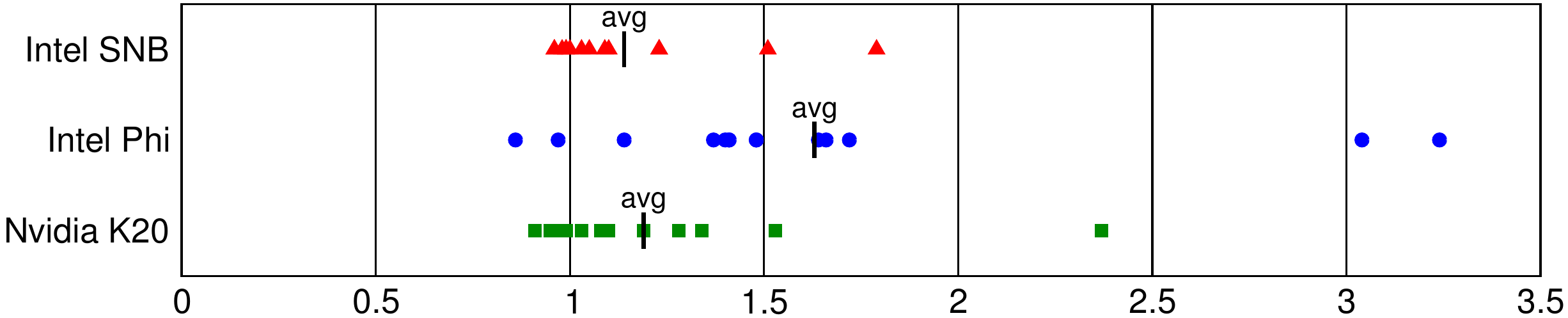}
\caption{Relative performance benefit of the unified SELL-32-$\sigma$ format over the vendor supplied library SpMVM performance for all non-pathological test cases}
\label{fig:scatter}
\end{figure}
For the memory-bound cases in
\autoref{fig:perf_mem} a rather constant, high performance level
can be achieved for all ``large-$N_\mathrm{nzr}$'' matrices
($N_\mathrm{nzr}\gg 12$; see \autoref{sect:perfmodels}), e.g., RM07R
($N_\mathrm{nzr}=98.16$) and ML\_Geer ($N_\mathrm{nzr}=73.72$). This
is in good agreement with our performance model: The maximum performance
for this scenario on all compute devices can be estimated
by setting $\beta=1$ in \autoref{best_veccache}. Choosing the best
memory bandwidth measurement from \autoref{tab:architectures} for each
architecture, we find $\bar{P}= 7.2\,\GFS$ (Intel SNB),
$\bar{P}=27.5\,\GFS$ (Intel Phi), and $\bar{P}=25.2\,\GFS$ 
(\nvidia\ K20). For most matrices with $N_{\mathrm{nzr}}> 50$ 
the \nvidia\ K20 and Intel Phi achieve around $80\%$ of this theoretical
limit and the Intel SNB gets more than $90\%$. For
``low-$N_\mathrm{nzr}$'' matrices, e.g.,  Hamrle3 and kkt\_power,
performance drops by a factor of roughly two or more for all
architectures. This is also in line with the performance model, which
will be discussed below in \autoref{sect:perf_mod_val}. Of course,
these performance models do not hold for cache bound matrices, as can be seen
in \autoref{fig:perf_cache}.

The Intel Phi delivers a disappointing performance on kkt\_power as
compared to the other architectures for all data layouts. 
A high coefficient of variation (cf. \autoref{eq:cv}) of 1.05 implies
that performance for this matrix suffers from load imbalance.

Worst accelerator performance is found for rail4284, where the small
number of rows ($N=4,284$) does not provide enough parallelism for both
architectures. Even for the simple copy benchmark (see
\autoref{sec:hardware}) the \nvidia\ K20 requires more than $10^4$
threads to hide memory and execution unit latencies. In these problematic cases,
issuing multiple
threads per row ($T$) on the GPGPU increases parallelism and thus performance. This has
been demonstrated by the related ``Sliced ELLR-T'' format
\cite{dziekonski11} and can be easily implemented in \SELLCS\ as well.
Using $T=32$, the performance on \nvidia\ K20 can be increased by a factor of nine
(from 0.5 \GFS to 4.5 \GFS). A reason for this still very low performance is the 
large coefficient of variation of 1.6 for this matrix.

Another poor-performing matrix on accelerators is webbase-1M. Its low
$N_\mathrm{nzr}$ in combination with a very small $\beta$ value indicate a highly irregular access pattern.
In addition, this matrix has an extremely large coefficient of variation of 8.16
which signifies a large likelihood of load imbalance. The large $\zeta$ is due to the fact
that this matrix contains a single row which is fully occupied with $1,000,005$ non-zero elements
while the average row length is merely 3.11.
Also Choi et al. \cite{choi10}
found very low performance levels on previous \nvidia\ GPGPU generations
for this matrix and omitted it in their further discussion. 
Due to the smaller number of threads the impact of load imbalance is much smaller
on Intel SNB. Hence, the performance on this architecture meets the expectation from the performance model as
described in \autoref{sect:perf_mod_val}.

A small row count is also the main reason for the low performance of Intel
Phi and \nvidia\ K20 in the dense2 case (a $2000\times 2000$ dense
matrix). Note that with a sufficiently large (i.e., $8000\times 8000$) dense 
matrix (as used in \cite{Liu:2013:ESM:2464996.2465013}),
much higher performance can be reached on all architectures.
Specifically, on Intel SNB we reach $7.3\,\GFS$, on Intel Phi we get $23\,\GFS$ 
(with ``\verb.STATIC.'' scheduling), and on \nvidia\ K20 we see $17\,\GFS$ 
($31\,\GFS$ with $T=16$).

Load imbalance and a small $\beta$ are also the reasons for the comparatively low accelerator performance for
the scircuit matrix despite the small matrix size. 
In this case, 94 \% of rows have less than 10 entries but the longest row has 353 entries.
This leads to a dominance of the HYB data format over \SELLCS\ for the \nvidia\ K20.

To conclude the performance discussion we are presenting \autoref{fig:scatter} which represents the relative
performance benefit of the \SELLCS\ over the baseline for all non-pathological test matrices.

\subsection{Performance model validation}\label{sect:perf_mod_val}

\begin{figure}[tbp]
\includegraphics[width=0.6\columnwidth]{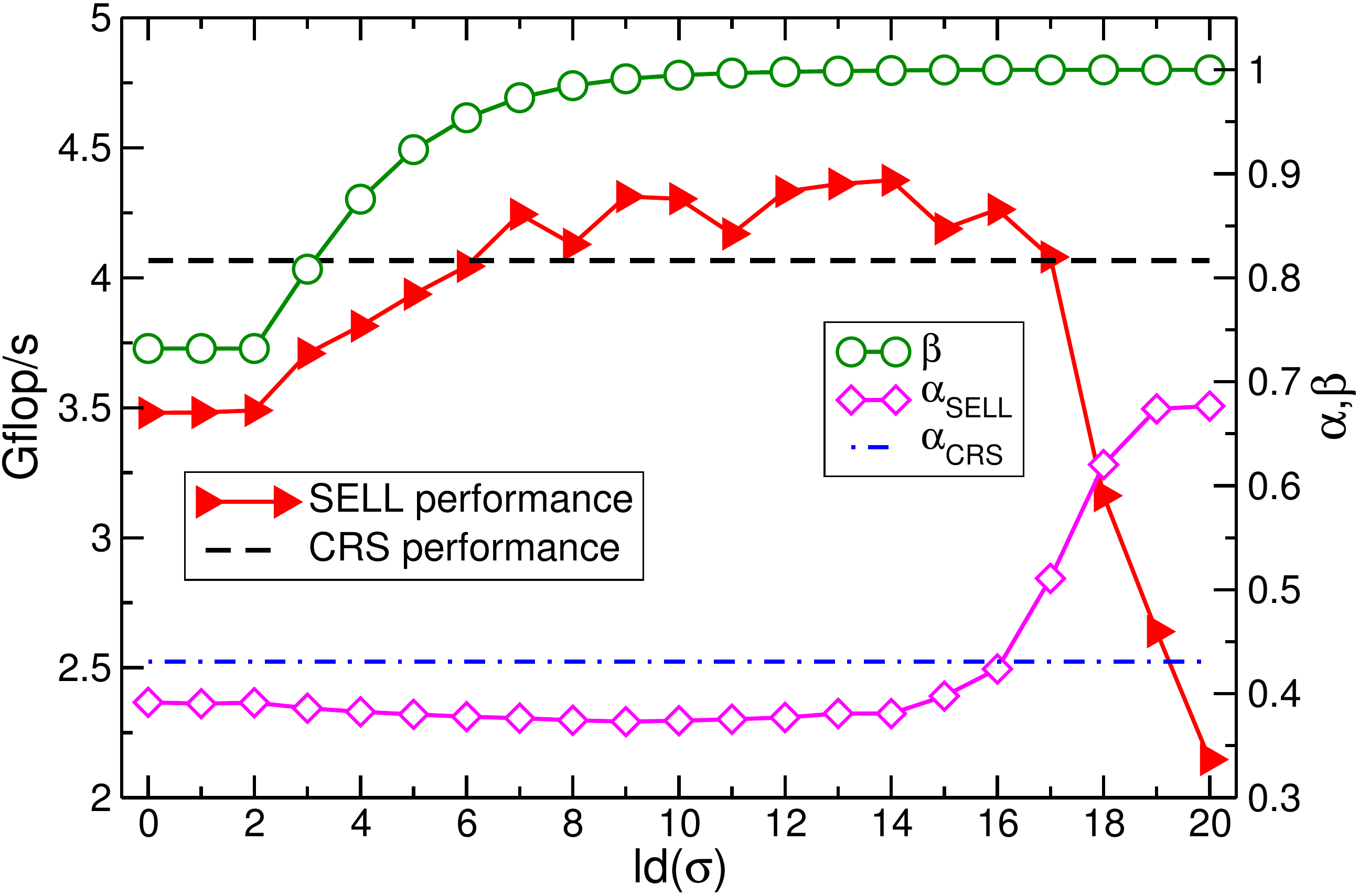}\hfill
\begin{minipage}[b]{.38\textwidth}
\vfill
\caption{Impact of sorting scope size with the SELL-4-$\sigma$ format
  on performance (left ordinate), $\alpha$, and $\beta$ (both right
  ordinate) on Intel SNB for the kkt\_power matrix.}
\label{fig:alphabeta}
\end{minipage}
\end{figure}
The performance characteristics of spMVM with \SELLCS\ are the result of a
subtle interplay between the sorting scope $\sigma$, the RHS re-use
factor $\alpha$, and the \cocc\ $\beta$.
\autoref{fig:alphabeta} shows the impact of varying $\sigma$ on
$\alpha$, $\beta$, and the performance using the kkt\_power matrix on the Intel
SNB. In principle the LLC of this processor is large enough to hold the RHS
vector in this case and we should expect
$\alpha=1/N_\mathrm{nzr}=0.14$ to be constant. However, the RHS
data set of $16$\,\MiB\ would take up 80\% of the LLC;
competition with matrix and LHS data causes extra evictions
in this case, and the RHS must be loaded more than once. Moreover, the memory
bandwidth drawn by the spMVM (measured using hardware performance
counters) is always subject to some fluctuations/inaccuracies. Thus
determining $\alpha$ via \autoref{eq:sellalpha} provides a qualitative
rather than an exact quantitative picture.

Without sorting ($\sigma=1$), \SELLCS\ is slower than the CRS
due to the low $\beta=0.73$. Additionally,
$\alpha_{\mathrm{SELL}}$ is roughly equal to $\alpha_{\mathrm{CRS}}$.
This meets our expectations, because the RHS re-use factor should not
change between the two matrix formats qualitatively. Due to $C=4$ (AVX), $\beta$ is
constant for $1 \leq \sigma \leq 4$.  Also, neither the \SELLCS\
performance nor $\alpha$ change within this $\sigma$ range, which 
shows that sorting with a small scope does not disturb the RHS access
for this particular matrix; such a behavior cannot be expected in the general 
case, however.  When going to larger sorting scopes, we can observe
that $\beta$ converges to one, as expected. At the same time, \SELLCS\
performance increases and exceeds the CRS performance at $\sigma\approx 128$. 
Simultaneously, $\alpha_{\mathrm{SELL}}$ stays on the same
level until it increases sharply starting at $\sigma\approx2^{15}$. 
Hence, sorting the matrix rows with scopes smaller than this value
does not negatively affect the RHS access pattern. Beyond this
``threshold'' the increase of $\alpha_{\mathrm{SELL}}$ is accompanied by
a drop in \SELLCS\ performance.

\begin{figure}[tbp]
\includegraphics[width=0.6\columnwidth]{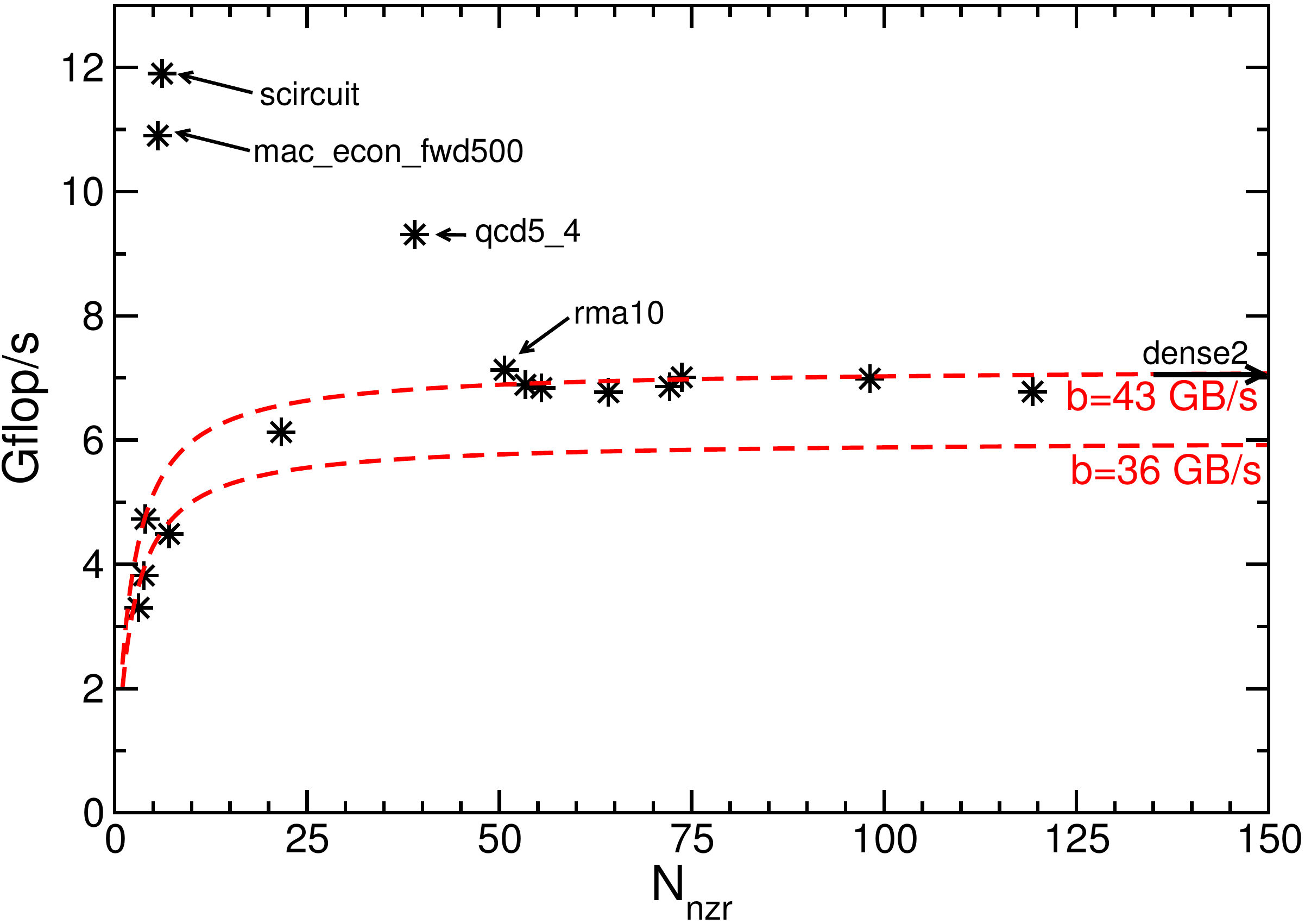}\hfill
\begin{minipage}[b]{.38\textwidth}
\vfill
\caption{Performance of SELL-4-opt on Intel SNB for all square test matrices
  depending on $N_\mathrm{nzr}$. 
  Dashed lines represent the prediction of the performance
  model (\ref{roof_alpha}) using the bandwidth range given in
  \autoref{tab:architectures}.}
\label{fig:snbperfm}
\end{minipage}
\end{figure}

Finally, we validate that the SIMD-vectorized (GPGPU-friendly)
\SELLCS\ format is able to attain high performance on Intel SNB
for all memory-bound matrices used in our work. According to the
discussion in \autoref{sect:perfmodels}, \autoref{roof_alpha} with
$\beta=1$ is an upper performance limit on Intel SNB, where the basic
model assumption holds that the complete RHS vector can stay in cache
during a single spMVM. Using the achievable bandwidth numbers from
\autoref{tab:architectures},
a maximum and minimum expected performance range as a function of
$N_\mathrm{nzr}$ is given in \autoref{fig:snbperfm} along with the
SELL-4-opt performance numbers for all square matrices. We find very good agreement
between the measurements and the model for all memory-bound cases. In
particular, \autoref{fig:snbperfm} demonstrates that the low
performance for webbase-1M ($N_\mathrm{nzr}=3.11$) and Hamrle3
($N_\mathrm{nzr}=3.81$), representing the two left-most stars in
\autoref{fig:snbperfm}, is caused by their short rows and can not be
improved substantially by a different data format. The matrices that
exceed the performance model have a memory footprint which easily fits
into the LLC of the Intel SNB (scircuit and mac\_econ\_fwd500) or are 
close to it (qcd5\_4 and rma10). 

\section{Conclusions and outlook}

We have motivated the need for a unified storage format for general
sparse matrices on modern compute devices. \SELLCS, which is ``Sliced ELLPACK''
combined with SIMD vectorization, was identified as the
ideal candidate. Although originally
designed for GPGPUs, \SELLCS\ is well suited for \emph{all} modern, 
threaded architectures with SIMD/SIMT execution such as the Intel
Xeon Sandy Bridge, Intel Xeon Phi, and \nvidia\ Kepler. Moreover,
\SELLCS\ is applicable to to a wide range of matrix types. This is a 
major step towards performance portability of spMVM kernels,
and enables the
possibility of running spMVM-based algorithms on heterogeneous compute
systems with the advantage of storing the matrix in a single format.
For most matrices investigated there is no significant loss of performance
compared to hardware-specific formats. For Intel Xeon Phi,
\SELLCS\ outperforms  CRS on most
matrices significantly, and may set a new standard sparse matrix data
format on this architecture. By construction, \SELLCS\ is ready to
exploit future architectures which are expected to deliver performance
mainly through wide SIMD/SIMT execution.  Thus, \SELLCS\ allows the
straightforward, efficient use of hybrid programming models like
OpenACC, OpenCL, or offload programming with the Xeon Phi, and is
expected to be easily portable to future computer architectures.

In future work we will address the challenge of increasing accelerator
performance of \SELLCS\ for matrices with small row count or many
non-zeros per row by adopting the well-known GPGPU strategy of running
multiple threads per row. A next logical step would be the
implementation of an MPI-enabled spMVM based on \SELLCS\ for use on
hybrid compute clusters. Additionally, the (automatic) selection of
tuning parameters like the sorting scope or the number of threads
covering a chunk should be considered.  Another question which has to
be answered is whether and to what extent \SELLCS\ is suited for other
numerical kernels besides spMVM.

\section{Acknowledgments}

We are indebted to Intel Germany and \nvidia\ for providing test
systems for benchmarking. This work was supported (in part) by the
German Research Foundation (DFG) through the Priority Programme 1648
``Software for Exascale Computing'' (SPPEXA) under project ESSEX, by
the Competence Network for Scientific High Performance Computing in
Bavaria (KONWIHR) under project HQS@HPC, and by the U.S. Department of
Energy (DOE).

\bibliographystyle{drgh}
\bibliography{sigproc}

\newpage
\begin{appendix}

\section{Description of the corner case benchmark matrices}\label{sect:cornercases}

\renewcommand{\labelenumi}{\textbf{(\alph{enumi})}} 

We conduct the detailed performance analysis of various storage 
formats based on four matrices from The
University of Florida Sparse Matrix
Collection\footnote{\url{http://www.cise.ufl.edu/research/sparse/matrices}}. 
The following descriptions are taken from the same source:
\begin{enumerate}

\begin{minipage}{.55\columnwidth}
\item \textbf{RM07R}\\ 
This matrix emerges from a CFD finite-volume
  discretization and represents a 3D viscous case with ``frozen''
  turbulence.
\end{minipage}
\hfill
\begin{minipage}{.2\columnwidth}
\centering
\includegraphics[width=\textwidth]{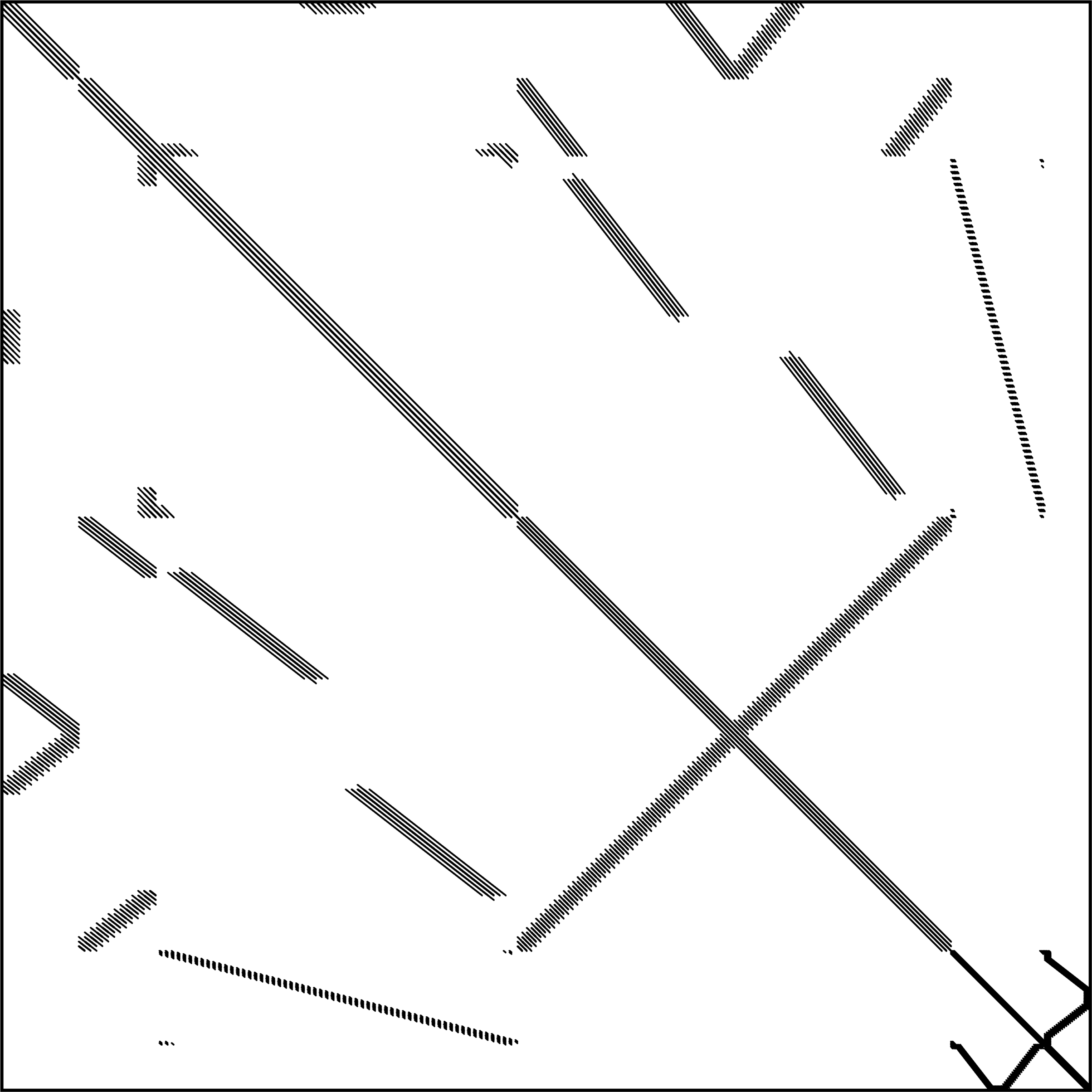}
\end{minipage}

\begin{minipage}{.55\columnwidth}
\item \textbf{kkt\_power} \\
This matrix arises from a non-linear optimization (Karush-Kuhn-Tucker) for finding the optimal power flow.
\end{minipage}
\hfill
\begin{minipage}{.2\columnwidth}
\centering
\includegraphics[width=\textwidth]{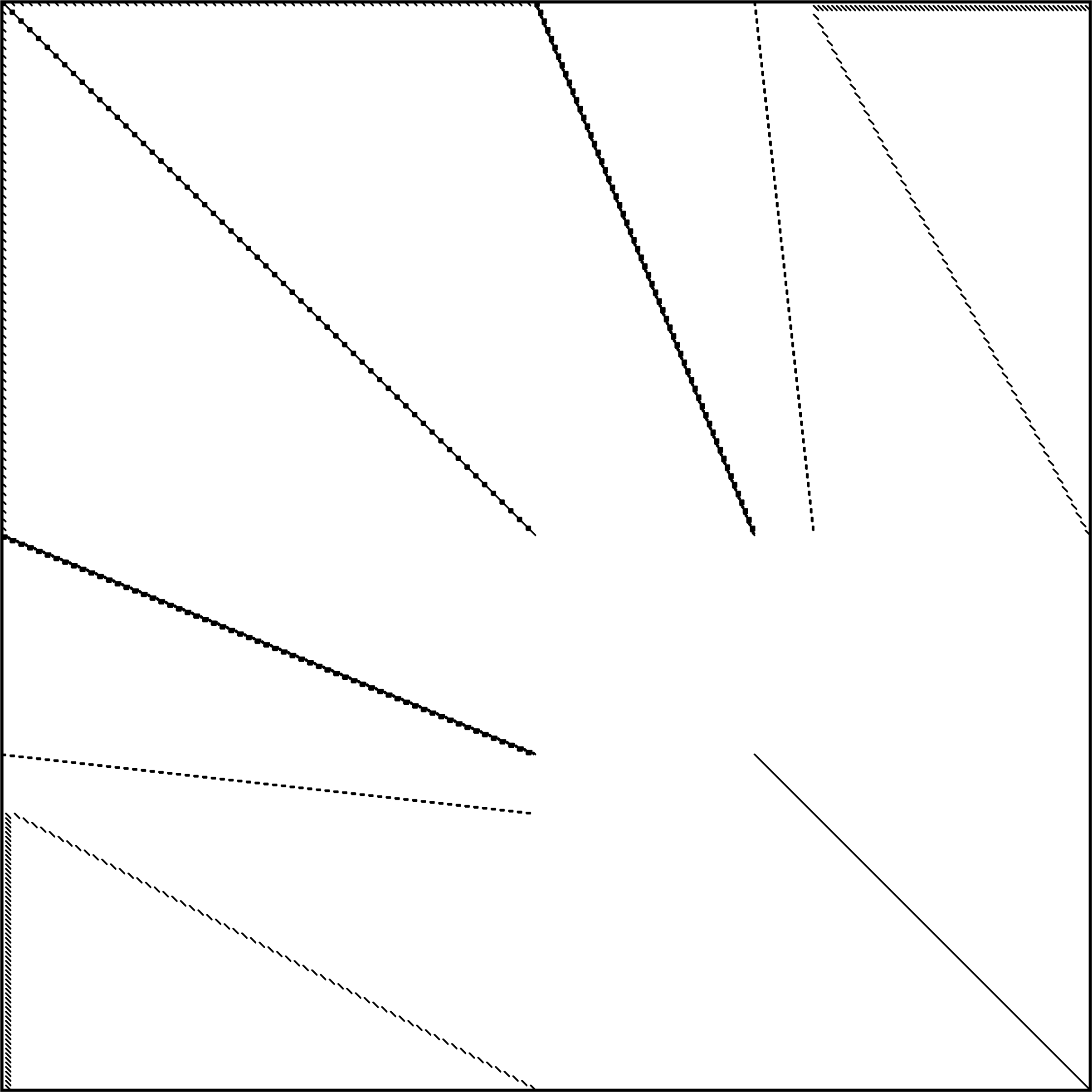}
\end{minipage}

\begin{minipage}{.55\columnwidth}
\item \textbf{Hamrle3} \\
This matrix originates from a very large electrical network simulation.
\end{minipage}
\hfill
\begin{minipage}{.2\columnwidth}
\centering
\includegraphics[width=\textwidth]{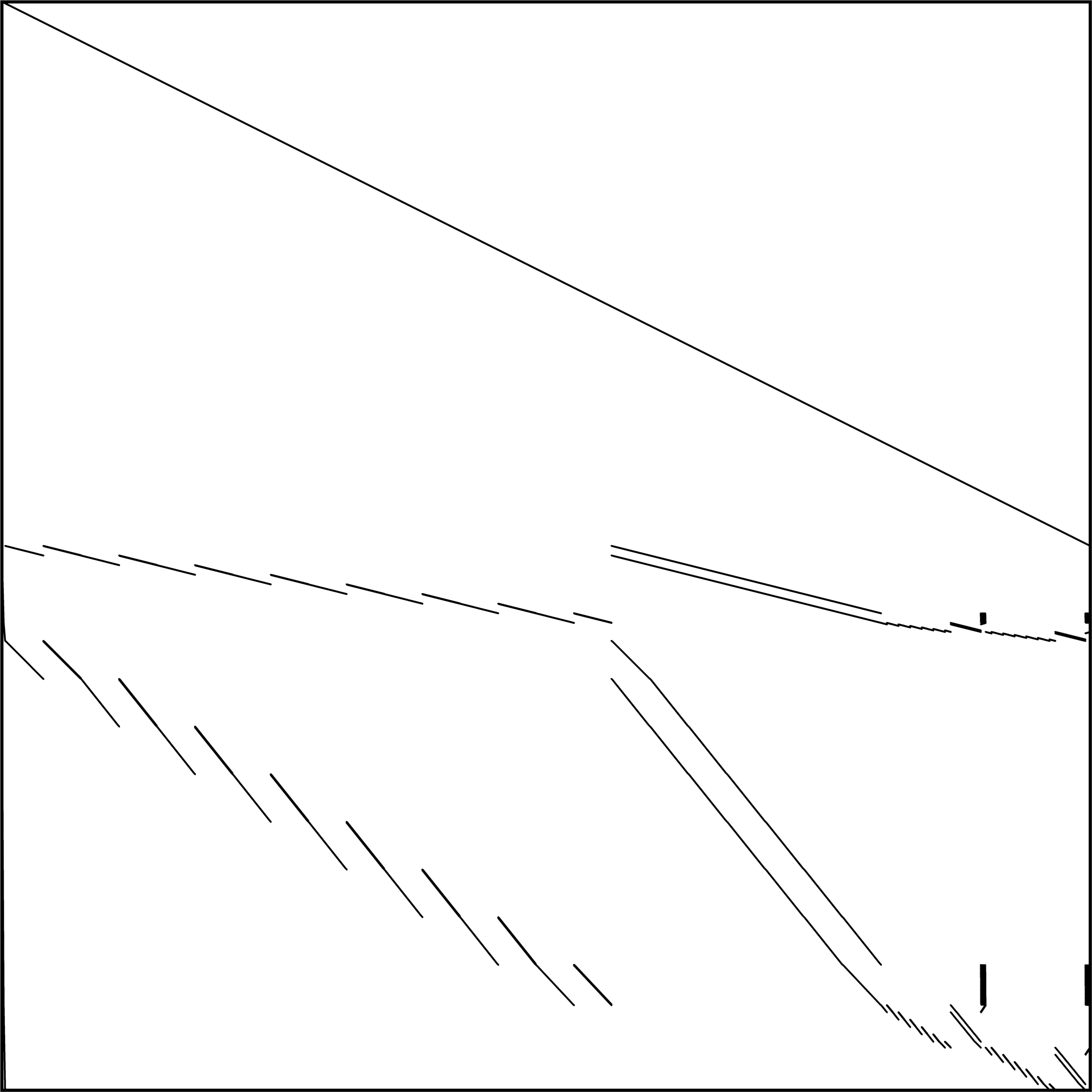}
\end{minipage}

\begin{minipage}{.55\columnwidth}
\item \textbf{ML\_Geer} \\ 
This matrix has been obtained to find the
  deformed configuration of an axial-symmetric porous medium subject
  to a pore-pressure drawdown through a meshless Petrov-Galerkin
  discretization.
\end{minipage}
\hfill
\begin{minipage}{.2\columnwidth}
\centering
\includegraphics[width=\textwidth]{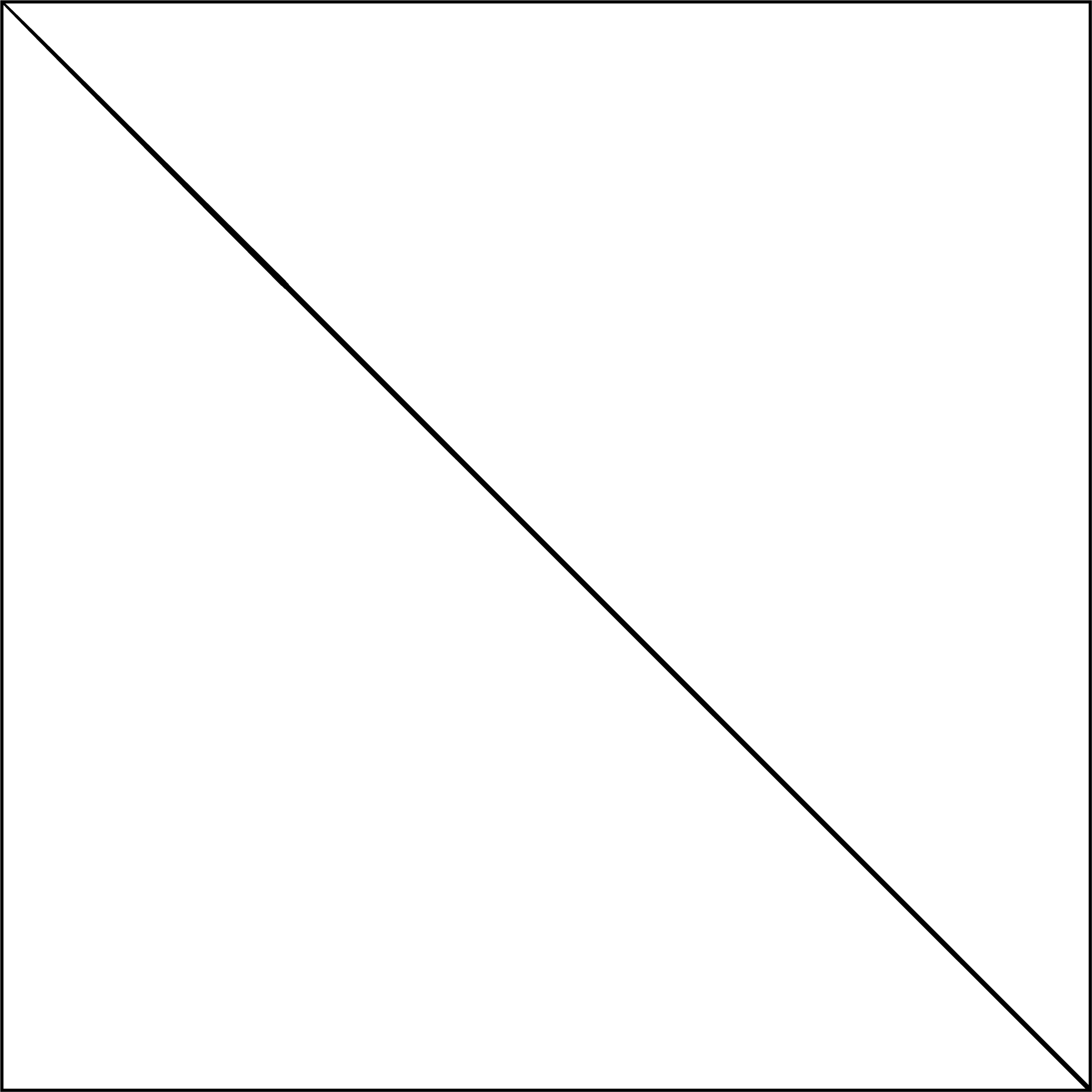}
\end{minipage}
\end{enumerate}
\newpage

\section{Code listings}

\lstset{caption={\SELLCS\ kernel for 64-bit values and 32-bit indices
    ($C=16$) implemented using Intel MIC intrinsics}\label{listing:sell_micintr}}
\begin{lstlisting}
int c, j, offs;
__m512d tmp1, tmp2, val, rhs;
__m512i idx;

#pragma omp parallel for schedule(runtime) private(j,offs,tmp1,tmp2,val,rhs,idx)
for (c=0; c<nRowsPadded>>4; c++) 
{ // loop over chunks
  tmp1 = _mm512_load_pd(&_lhs[c<<4]  );    // load 8 LHS values
  tmp2 = _mm512_load_pd(&_lhs[c<<4+8]);    // load next 8 LHS values
  offs = cs[c]; // the initial offset is the start of this chunk

  for (j=0; j<cl[c]; j++) 
  { // loop inside chunk from 0 to the length of the chunk
    val = _mm512_load_pd(&_val[offs]);       // load 8 matrix values
    idx = _mm512_load_epi32(&_col[offs]);    // load 16 indices
    rhs = _mm512_i32logather_pd(idx,_rhs,8); // gather RHS using lower 8 indices
    tmp1= _mm512_add_pd(tmp1,_mm512_mul_pd(val,rhs)); // multiply & accumulate
    offs+= 8;

    val = _mm512_load_pd(&_val[offs]);      // load next 8 matrix values
    idx = _mm512_permute4f128_epi32(idx,_MM_PERM_BADC); // lo <-> hi idx
    rhs = _mm512_i32logather_pd(idx,_rhs,8);// gather rhs lower 8 indices
    tmp2= _mm512_add_pd(tmp2,_mm512_mul_pd(val,rhs)); // multiply & accumulate
    offs += 8;
  }

  _mm512_store_pd(&_lhs[c<<4]  , tmp1);    // store 8 LHS values
  _mm512_store_pd(&_lhs[c<<4+8], tmp2);    // store next 8 LHS values
}
\end{lstlisting}

\lstset{caption={\SELLCS\ kernel for 64-bit values and 32-bit indices
    ($C=4$) implemented using Intel AVX intrinsics}\label{listing:sell_avxintr}}
\begin{lstlisting}
int c, j, offs;
__m256d tmp, val, rhs;
__m128d rhstmp;

#pragma omp parallel for schedule(runtime) private(j,offs,tmp,val,rhstmp)
for (c=0; c<nRowsPadded>>2; c++) 
{ // loop over chunks
  tmp = _mm256_load_pd(&_lhs[c<<2]); // load 4 LHS values
  offs = cs[c]; // the initial offset is the start of this chunk

  for (j=0; j<cl[c]; j++) 
  { // loop inside chunk from 0 to the length of the chunk
    val    = _mm256_load_pd(&_val[offs]);               // load 4 matrix values
    rhstmp = _mm_loadl_pd(rhstmp,&_rhs[_col[offs++]]);  // load 1st RHS value
    rhstmp = _mm_loadh_pd(rhstmp,&_rhs[_col[offs++]]);  // load 2nd RHS value
    rhs    = _mm256_insertf128_pd(rhs,rhstmp,0);        // insert lo part of RHS
    rhstmp = _mm_loadl_pd(rhstmp,&_rhs[_col[offs++]]);  // load 3rd RHS value
    rhstmp = _mm_loadh_pd(rhstmp,&_rhs[_col[offs++]]);  // load 4th RHS value
    rhs    = _mm256_insertf128_pd(rhs,rhstmp,1);        // insert hi part of RHS
    tmp    = _mm256_add_pd(tmp,_mm256_mul_pd(val,rhs)); // accumulate
  }

  _mm256_store_pd(&_lhs[c<<2], tmp); // store 4 LHS values
}
\end{lstlisting}

\end{appendix}

\end{document}